\documentclass[12pt]{article}
\usepackage{graphicx}
\usepackage{hyperref}
\usepackage{amsmath}
\usepackage{amscd,amssymb}

\newtheorem{Remark}[equation]{Remark}

\newtheorem{Statement}[equation]{Statement}

\newcommand{\fz}{\mathfrak{z}}

\def\cR{{\mathcal R}}

\def\cJ{{\mathcal J}}

\def\cC{{\mathcal C}}
\def\cK{{\mathcal K}}

\def\cH{{\mathcal H}}

\def\cL{{\mathcal L}}

\def\cY{{\mathcal Y}}

\def\cO{{\mathcal O}}

\def\mG{{\mathfrak G}}

\def\mg{{\mathfrak g}}

\def\ml{{\mathfrak l}}
\def\ms{{\mathfrak s}}
\def\mg{{\mathfrak g}}
\def\mf{{\mathfrak f}}

\newtheorem{theorem}{Theorem}[section]

\newtheorem{remark}{Remark}[section]

\def\e{{\,\rm e}\,}

\newcommand{\beq}{\begin{eqnarray}}
\newcommand{\eeq}{\end{eqnarray}}
\numberwithin{equation}{section}

\def\ch{{\rm ch}}

\newcommand{\xX}{\textsf{X}}
\newcommand{\MM}{\textsf{M}}

\begin{document}


\begin{center}
{\large\bf $S$-Functions, Spectral Functions of Hyperbolic Geometry, and Vertex Operators
with Applications to Structure for Weyl and Orthogonal Group Invariants}
\end{center}

\vspace{0.1in}

\begin{center}
{\large
A. A. Bytsenko $^{(a)}$
\footnote{E-mail: aabyts@gmail.com} and
M. Chaichian $^{(b)}$
\footnote{E-mail: masud.chaichian@helsinki.fi}}

\vspace{5mm}

$^{(a)}$
{\it
Departamento de F\'{\i}sica, Universidade Estadual de
Londrina\\ Caixa Postal 6001,
Londrina-Paran\'a, Brazil}

\vspace{0.2cm}
$^{(b)}$
{\it
Department of Physics, University of Helsinki\\
P.O. Box 64, FI-00014 Helsinki, Finland}


\end{center}

\vspace{0.1in}

\begin{abstract}
In this paper we analyze the quantum homological invariants (the Poincar\'e polynomials of
the $\mathfrak{sl}_N$ link homology). In the case
when the dimensions of homologies of appropriate topological spaces are precisely known, the
procedure of the calculation of the Kovanov-Rozansky type homology, based on the
Euler-Poincar\'e formula can be appreciably simplified. We express the formal character of
the  irreducible tensor representation of the classical groups  in terms
of the symmetric and spectral functions of hyperbolic geometry.
On the basis of Labastida-Mari\~{n}o-Ooguri-Vafa conjecture, we derive a representation
of the Chern-Simons partition function in the  form of an infinite product in terms of the Ruelle
spectral functions (the cases of a knot, unknot, and links have been considered).
We also derive an infinite-product formula for the orthogonal Chern-Simons partition
functions and analyze the singularities and the symmetry properties of the
infinite-product structures.
\end{abstract}

\vspace{0.1in}

\begin{flushleft}
PACS \, 02.10.Kn, 02.20.Uw, 04.62.+v
\\
\vspace{0.3in}
\end{flushleft}


\begin{flushright}
{\bf\emph{Dedicated to the memory of  our friend and colleague, Petr P. Kulish}}
\end{flushright}
\newpage

\tableofcontents


\section{Introduction}

{\bf Graded Poincar\'e polynomials; infinite-dimensional algebras.}
The aim of this paper is to exploit and emphasis the structure of
quantum group invariants. Recall some recent activities related to quantum invariants and
the homological invariants of the Hopf link.

Note a certain importance in many diverse areas of mathematics and physics a class of
infinite-dimensional algebras, in particular (affine) Kac-Moody algebras, which
has been introduced in the late 1960's.
Unlike of the finite-dimensional case (where simple Lie algebras can be realized in terms of
a finite number of bosonic/fermionic {\it modes}), infinite-dimensional Kac-Moody algebras
have various vertex operator realizations (in terms of a finite number of bosonic {\it free
fields}, the modes of which generate a Heisenberg algebra).
Generally speaking, all simple (twisted and untwisted) Kac-Moody algebras can be embedded
in the infinite-dimensional algebra ${\mathfrak g}{\mathfrak l}(\infty)$ of infinite
matrices with a finite number of non-zero entries (see Sect. \ref{Hopf}), which has
a simple realization in terms of generators of a Clifford algebra.

Virasoro algebra as an another type of infinite-dimensional algebra arises in different
areas of physics (see Sect \ref{Hopf} for details); this is the algebra of conformal
transformations in two-dimensions. The operator algebra structure of two-dimensional
conformally-invariant quantum field theories is determined by the representation
theory of the Virasoro algebra. The Ramond and Neveu-Schwarz superalgebras
(or the $N=1$ superconformal algebras) are supersymmetric extensions of the Virasoro
algebra.

We also mention the quantum affine algebras, which are $q$-deformations of Kac-Moody
algebras. By analogy with the undeformed case, vertex
operator realizations, initially for level one representations and then
for arbitrary high level representations, were constructed.

Symmetric functions (or $S$-functions) with it connection to replicated plethisms,
has been involved in the applications to certain infinite-dimensional Lie algebras
(in particular the quantum affine algebras $U_q({\mathfrak g}{\mathfrak l}_N)$) and
generating functions of quantum ${\mathfrak g}{\mathfrak l}_N$ invariants.
The $S$-functions were first studied by Jacobi and have been generalized in various
applications in physics and mathematics. There are numerous generalizations of $S$-functions,
and among them Jack symmetric functions. Jack symmetric functions are just a special limit of
a generalized Hall-Littlewood function 
considered by Macdonald \cite{Macdonald}.
Macdonald's polynomials can be formulated as the trace of an
interwiner (algebra homomorphysm) of modules over the quantum group \cite{Etingof}.

{\bf Quantum group invariants; finite-dimensional algebras.}
Another type of algebra, which has had a wide variety of applications in
physics, is the so-called quantum group; this may be regarded as a deformation, depending
on a parameter $q$, of the universal enveloping algebra of a semi-simple Lie algebra.
Thus they are not finite-dimensional algebras, but are finitely generated.
These algebras were first constructed by Kulish and Reshetikhin \cite{KR} and as a Hopf
algebra by Sklyanin \cite{Sklianin}.
Their representation theory for $q$ not a root of unity was found to be simliar to the
corresponding semi-simple Lie algebra, complications arise when $q$ is a root of unity due to
the fact that the centre of the algebra becomes larger.
Quantum groups are an example of quasi-triangular Hopf algebras and as such,
for each quantum group there exists a universal $R$-matrix which intertwines with the
action of the coproduct.

By exploiting calculations for oriented and unoriented links the two-variable HOMFLY and
one-variable Kauffman polynomials have been analyzed in \cite{HOMFLY,PT} and \cite{Kauffman1}
respectively. Note that HOMFLY polynomial can be generalized to Alexander and Jones polynomial;
the two-variable Kauffman polynomial has been also introduced in \cite{Kauffman2} by
generalization of the Jones polynomial.
A polynomial invariant of oriented knots has been discovered in \cite{Jones1,Jones2}.
A quantum field theory interpretation of the Jones polynomial by the Chern-Simons path integral
method has been given by Witten \cite{Witten}, who also predicted the existence of 3-manifold
quantum invariants.

A construction of three-manifold invariants via a track of quantum universal enveloping algebra
(quantum group) $U_{q}({\mathfrak s}{\mathfrak l}_{2})$ at roots of unity has been discovered by
Reshetikhin and Turaev in \cite{RT1, RT2}; it leads to the {\it colored} version of classical
HOMFLY and Kauffman polynomial invariants. {\it Color} in this context means the representation
of quantum groups, and colored HOMFLY polynomial is a special linear quantum group invariants,
the quantum group of $A_{n}$ type, whereas colored Kauffman polynomial invariants is a
quantum group invariants of $B_{n}$, $C_{n}$ and $D_{n}$ type. These achievements actually
give a unified understanding of the quantum group invariants of links. Generalization to
affine Macdonald's polynomials by considering homomorphisms of the affine quantum group
$U_q({\mathfrak g}{\mathfrak l}_N)$ has been considered in \cite{etingof:kirillov:1995a}.

In articles \cite{LaM,LMV,OV} a conjectural description of relationship between
reformulated invariants of colored HOMFLY link has been proposed; we will refer to this
proposal as Labastida-Mari\~{n}o-Ooguri-Vafa (LMOV) conjecture. Later this conjecture was proved
in \cite{Liu2}. The LMOV conjecture can be expressed by using mathematical language, theory
of irreducible representation of quantum groups. The physics background can be addressed to
works in late 1970's on large $N$ expansion of $U(N)$ gauge field theories.
In this connection, the exact result for closed topological string theory on the resolved
conifold, dual to the $U(N)$ Chern-Simons theory on $S^{3}$, has been described by Gopakumar
and Vafa \cite{GV}. The Gromov-Witten theory of the resolved conifold corresponds to the
Chern-Simons theory of an unknot, while the LMOV conjecture considers the case when the link
or knot is nontrivial and the corresponding Wilson loop expectation values, in fact colored
HOMFLY polynomial of the link. Therefore the LMOV conjecture could be viewed as a counterpart
of Gopakumar-Vafa conjecture.
It is important that the LMOV conjecture predicts an intrinsic symmetry of series in
$q-q^{-1}$ about reformulated invariants of the colored HOMFLY polynomial as well as hidden
integrality encoded in the colored HOMFLY polynomial.

The orthogonal LMOV conjecture was formulated using the colored Kauffman solely in \cite{C-C},
and the relation between the colored HOMFLY polynomials and the colored Kauffman polynomials
in \cite{Marino} correspondingly.
In recent articles \cite{LiuPeng,ChenLiu} a new structure of the colored HOMFLY polynomial has
been analyzed by means of the Chern-Simons partition function, which appear in the LMOV
conjecture as an infinite product, and indicates some potential modularity of this partition
function.

{\bf The organization of the paper and our key results.}

-- The quantum homological invariants we consider in Sect. \ref{Poincare}. We analyze
Poincar\'e polynomial of the $\mathfrak{sl}_N$ link homology
$\overline{\mathcal P}_{\mathfrak{sl}_N; \lambda, \mu} ({q},{t})$. We show that in the case
when dimensions of homologies of appropriate topological spaces are precisely known the
procedure of the calculation of the Kovanov-Rozansky type homology, based on the
Euler-Poincar\'e formula, can be appreciably simplified.

-- In Sect. \ref{Hopf} we exploit the Hopf algebra structure of the ring $\Lambda(X)$ of
$S$-functions of the independent variables $(x_1, x_2, \cdots)$ (finite or countably infinite
in number), that constitute the alphabet $X$, mainly following to notation and discussion of
important article \cite{Fauser10}. After some notational preliminaries
we discuss algebraic properties of the ring $\Lambda(X)$ in Sect. \ref{Ring}.
Then in Sect. \ref{Plethysm} we define plethisms (Schur function plethism, scalar and inner
(co)products), the notation for the mutually inverse pair and the inner products is:
$M(t;X) = \prod_{i\geq 1}(1-tx_i)^{-1},\, L(t; X) = \prod_{i\geq 1}(1-tx_i)$ and
$M(XY) = \prod_{i,j}(1-x_iy_j^{-1})$, $L(XY)= \prod_{i,j}(1-x_iy_j)$; the Cauchy kernel,
$M(XY)$, is a dual version of the Schur-Hall scalar product.
The next important result realizes more plethisms associated with multipartite generating
functions, Sect. \ref{Mult}. It is obtained through the use of one of the restricted
specializations. In addition, we use the Bell polynomials in multipartite function problem
with it connection to the spectral functions of hyperbolic 3-geometry. We conclude with
some examples of hierarchy which can be treated as a product of copies, each of them
corresponds to a free two-dimansional CFT (Eq. (\ref{CFT_2})). The relevant formulas of
this section expressed in terms of so-called Ruelle (Patterson-Selberg) spectral zeta-functions
(\ref{R1}) -- (\ref{RU2}) are very efficient.

-- The characters of the orthogonal and symplectic groups have been
found by Schur \cite{Schur,schur:1924a} and Weyl \cite{Weyl} respectively.
The method has been used is transcendental (it depends on integration over the group manifold).
However the appropriate characters may also be obtained by algebraic methods
\cite{Littlewood44}: {\it ``This algebraic method would seem to offer a better prospect
of successful application to other restricted groups than the method of group integration.''}
Following \cite{Fauser10} we have used algebraic methods.
The Hopf algebra can be exploited in the determination of (sub)group branching rules and
the decomposition of tensor products. We use this analysis for vertex operator traces
in Sect. \ref{Character}.

-- The HOMFLY skein and the link invariants from vertex models we analyzed in Sect. \ref{HOMFLY}.
Then in Sect. \ref{Product} on the base of LMOV conjecture  we derive a new representation
of the Chern-Simons partition function in form of an infinite product in terms of Ruelle
spectral functions of hyperbolic geometry. In addition, we consider the case of a knot,
unknot, and links.   We discuss singularities and symmetry
properties of these infinite-product structures in Sect. \ref{Symmetry}.
Infinite-product formula for orthogonal Chern-Simons partition functions we derive
in Sect. \ref{Orthogonal}.

\section{Poincar\'e polynomials of the homological invariants of the Hopf link}
\label{Poincare}

{\bf Polygraded algebras and polynomial invariants.} The
relationship between Lie algebras and combinatorial identities
(the famous Euler identity, as a particular famous example) was
first discovered by Macdonald. A general outline for proving
combinatorical identities is based on the Euler-Poincar\'e
formula. Let ${\mathfrak g}$ be a polygraded Lie algebra, $
{\mathfrak g} = \bigoplus_{\scriptstyle \lambda_1\geq 0, ...,
\lambda_k\geq 0 \atop\scriptstyle
\lambda_1+...+\lambda_k>0}{\mathfrak g}_{(\lambda_1, ...,
\lambda_k)}, $ satisfying the condition $ {\rm dim}\, {\mathfrak
g}_{(\lambda_1, ..., \lambda_k)} < \infty. $ For formal power
series in $q_1,...,q_k$, one can get the following identity (the
Euler-Poincar\'e formula):
\begin{equation}
{\mathcal P}_{\mathfrak g}(q) = \sum_{m,\lambda_1,...,\lambda_k}(-1)^m
q_1^{\lambda_1} \cdots q_k^{\lambda_k} {\rm dim}\, H_m^{(\lambda_1,...,\lambda_k)}
=
\prod_{n_1,...,n_k}\left(1-q_1^{n_1}\cdots q_k^{n_k}\right)^
{{\rm dim}\, {\mathfrak g}_{n_1, ..., n_k}}\,.
\label{poly}
\end{equation}

Interesting combinatorial identities may be obtained by applying (\ref{poly}) to graded algebras,
for example, to the subalgebras ${\mg}^A$ of Kac-Moody algebras.
From the point of view of the applications, homologies associated with subalgebras
$\mg = {\ms}{\ml}_{N} ({\mathbb C})$ are much more important than the homology of the current
algebra themselves, since they constitute the
thechnical basis of the proof of the combinatorial identities of Euler-Gauss-Jacobi-MacDonald.

{\bf Poincar\'e polynomial of the $\mathfrak{sl}_N$ link homology
$\overline{\mathcal P}_{\mathfrak{sl}_N; \lambda, \mu} ({q},{t}) $.}
It is known that the invariants of the colored Hopf link are
identified as topological open string amplitudes on the deformed conifold $T^*S^3$.
Indeed, the $U(N)$ Chern-Simons
theory is realized by topological strings on $T^*S^3$ with $N$ topological D-branes wrapping
on the base Lagrangian submanifold $S^3$.
On the other hand the Hopf link in $S^3$ consisting of two knots ${\mathcal K}_1$ and
${\mathcal K}_2$ can be introduced by a
pair of new D-branes wrapping on Lagrangian three-cycles $L_1$ and ${L}_2$
such that $S^3 \cap {L}_i = {\mathcal K}_i$ \cite{OV}. For this brane system the
topological open string amplitude
(rewritten in terms of symmetric functions and spectral Selberg-type functions) is supposed
to give the invariants of the Hopf link.
The representation (or coloring) attached to each knot ${\mathcal K}_i$ is related to the
boundary states of the open string ending
on ${L}_i$ by the Frobenius relation.
Because of geometric transition or the large $N$ duality \cite{GV,Vafa01,Ooguri02},
this brane configuration is mapped to the resolved conifold
${\mathcal O}(-1) \oplus {\mathcal O}(-1) \to
{\mathbb P}^1$. Eventually, the D-branes wrapping on $S^3$ disappear, but a pair of
Lagrangian D-branes remains as
a remnant of the Hopf link. The resulting D-brane system can be described in terms of
the toric diagram
and  the corresponding amplitude computed by the topological vertex method
\cite{Aganagis04,Aganagis05}.

Our interest is the superpolynomial which is a polynomial in $({a},{q},{t})\in{\mathbb C}^3$,
such that a specialization ${a} = {q}^N$ leads the Poincar\'e polynomial of the
$\mathfrak{sl}_N$ link homology
$\overline{\mathcal P}_{\mathfrak{sl}_N; \lambda, \mu} ({q},{t}) $,
which is a two parameter $({q},{t})$ version of the $\mathfrak{sl}_N$ link invariants.

In the case the coloring is the $N$ dimensional defining representation, it is called
the Khovanov-Rozansky
homology \cite{Khovanov}. In the case ${a}={q}^N$ with $N\in{\mathbb Z}_+$,
the superpolynomial of the homological invariants of the colored Hopf link reduces to
$
\sum_{i,j \in {\mathbb Z}} {q}^i {t}^j \dim {\mathcal H}_{i,j}^{\mathfrak{sl}_N;\lambda,\mu}
$
with certain doubly graded homology ${\mathcal H}_{i,j}^{\mathfrak{sl}_N;\lambda,\mu}$
\cite{Gukov10}
(by definition it should be a polynomial in ${q}$ and ${t}$ with non-negative integer
coefficients).
It has been argued \cite{GSV} that homological link invariants are related to a refinement
of the BPS state
counting in topological open string theory. The conjecture on homological link invariants
of the Hopf link
\cite{Gukov10} is based on this proposal. For the Hopf link $\mathcal L$ this means that
there is a doubly graded homology theory
${\mathcal H}_{i,j}^{\mathfrak{sl}_N; \lambda, \mu} ({\mathcal L})$
whose graded Poincar\'e polynomial is (Cf. Eq. (\ref{poly}))
\begin{equation}
\overline{\mathcal P}_{\mathfrak{sl}_N; \lambda, \mu} (q, t) = \sum_{i,j \in {\mathbb Z}}
{q}^i {t}^j \dim {\mathcal H}_{i,j}^{\mathfrak{sl}_N; \lambda, \mu} ({\mathcal L})~.
\label{super11}
\end{equation}
\begin{remark}
It is convenient for a computational reasons to investigate the effect of a relation on the
Poincar\'{e} series of a graded algebra.
Let ${\mathfrak g}_1$ and ${\mathfrak g}_2$ be two graded algebras. Suppose a basis for
${\mathfrak g}_1$ {\rm (}as a vector space{\rm )}
is $\{x_i\}_{i\in I}$,
while a basis for ${\mathfrak g}_2$ is $\{y_j\}_{j\in J}$. Then a basis for
${\mathfrak g}_1\otimes {\mathfrak g}_2$ {\rm (}as a vector space{\rm )} is
$\{x_i\otimes y_j\}_{i\in I,\,j\in J}$.
As a result, ${\mathcal P}_{{\mathfrak g}_1\otimes {\mathfrak g}_2}(q) =
{\mathcal P}_{{\mathfrak g}_1}(q){\mathcal P}_{{\mathfrak g}_2}(q)$.

Suppose that ${\mathfrak g}$ is a ring. We need the following definition {\rm \cite{Bott}:}
A sequence of elements $\{\gamma_j\}_{j=1}^r$ in
${\mathfrak g}$ is called a {\it regular sequence} if $\gamma_1$ is not a zero-divisor in
${\mathfrak g}$ and for each $j\geq 2$ the image of $\gamma_j$ in
${\mathfrak g}/(\gamma_1, \ldots, \gamma_{j-1})$ is not a zero-divisor.
Let, as before, ${\mathfrak g}$ be a graded algebra and $\gamma_1, \gamma_2, \ldots, \gamma_r$ a
regular sequence of homogeneous elements of degrees $n_1, n_2, \ldots, n_r$. We have {\rm (}see
for detail {\rm \cite{Bott})}
$
{\mathcal P}_{{\mathfrak g}/(\gamma_1, \ldots,\gamma_r)}(q) =
{\mathcal P}_{{\mathfrak g}}(q)(1-q^{n_1})\cdots (1-q^{n_r}).
$
\end{remark}
For more examples let us proceed to describing the properties of link homologies suggested by
the their relation to Hilbert spaces of BPS states \cite{GSV}.
Let ${\mathcal H}_{k,j}^{{\mathfrak{sl}_N; R_1, \ldots, R_\ell}}({\mathcal L})$
be the doubly-graded homology theory whose graded Euler characteristic
is the polynomial invariant $\overline P_{\mathfrak{sl}_N;R_1, \ldots, R_{\ell}} (q)$
(the {\it bar} means that this invariant is unnormalized invariant; its normalized version obtained by dividing by the invariant of the unknot)
\begin{equation}
\overline P_{\mathfrak{sl}_N;R_1, \ldots, R_{\ell}} (q) = \sum_{k,j \in \mathbb{Z}} (-1)^j q^k
\dim {\mathcal H}_{k,j}^{{\mathfrak{sl}_N; R_1, \ldots, R_\ell}}({\mathcal L}).
\label{pslncategor}
\end{equation}
Here $\mathcal L$ is an oriented link in $S^3$, we consider the Lie algebra ${\mathfrak g} =
\mathfrak{sl}_N$ (there is a natural generalization to other classical Lie algebras
$B$, $C$, and $D$ \cite{GSV})
and a link colored is given by a collection of representations $R_1, \ldots , R_\ell$ of
$\mathfrak{sl}_N$. The graded Poincar\'e polynomial has the form
\begin{equation}
\overline{{P}}_{\mathfrak{sl}_N;R_1, \ldots, R_{\ell}} (q,t) :=
\sum_{k,j \in \mathbb{Z}} q^k t^j
\dim {\mathcal H}_{k,j}^{{\mathfrak{sl}_N; R_1, \ldots, R_\ell}}({\mathcal L})\,.
\label{superpsln}
\end{equation}
By definition, it is a polynomial in $q^{\pm 1}$ and $t^{\pm 1}$
with integer non-negative coefficients. In addition,
evaluating (\ref{superpsln}) at $t=-1$ gives (\ref{pslncategor}).
In the case $R_a = \Box$ for all $a=1, \ldots, \ell$,
the homology ${\mathcal H}_{k,j}^{\mathfrak{sl}_N;
\Box, \ldots, \Box}({\mathcal L})$
is known as the Khovanov-Rozansky homology, ${{}_{(KR)}\overline{H}}_{k,j}^{N} ({\mathcal L})$.
The further physical interpretation of homological link invariants via
Hilbert spaces of BPS states leads to certain predictions regarding
the behavior of link homologies with rank $N$ (for more discussion see \cite{GSV,Gukov1}).

In the case when dimensions of homologies of appropriate topological spaces are precisely known the procedure of the calculation can be appreciably simplified. Such a situation happends, for example,  if asymptotic behavior the total dimension of ${\mathcal H}_{\ast, \ast}^{\mathfrak{sl}_N; R_1, \cdots, R_\ell} ({\mathfrak L})$ grows as \cite{Gukov10}:
$
{\rm dim}\,{\mathcal H}_{\ast, \ast}^{\mathfrak{sl}_N; R_1,
\cdots, R_\ell} ({\mathfrak L})\vert_{N\rightarrow\infty} \rightarrow N^d,\,
d = \sum_{j=1}^\ell {\rm dim}\,R_j\,.
$

\section{Hopf algebraic approach and group theory}
\label{Hopf}

Usually the Euler-Poincar\'e formula applies to chain complexes of finite dimensional
Lie algebras. In the infinite dimensional case matters can be fixed up by considering
polygraded Lie algebras. Thus the partition functions can indeed be converted into product
expressions.
The expression on the right-hand side of {\rm (\ref{poly})} looks like {\it counting}
the states in the Hilbert space of a second quantized theory.
Certain formulas for the partition functions or Poincar\'{e} polynomials
\begin{equation}
\prod_{n_1,...,n_k}\left(1-q_1^{n_1}\cdots q_k^{n_k}\right)^{{\rm dim}\,
{\mathfrak g}_{n_1, ..., n_k}},\,\,\,\,\,\,\,\,
\prod_{n_1,...,n_k}\left(1-q_1^{n_1}\cdots q_k^{n_k}\right)^
{{\rm rank}\, {\mathfrak g}_{n_1, ..., n_k}}
\label{Series}
\end{equation}
are associated with dimensions of homologies of
appropriate topological spaces and linked to generating functions
and elliptic genera. Note that
this conclusively explains the sequence of dimensions ({\it distinguished powers}) of the
simple Lie algebras.

Before examine a ring of formal power series (of type (\ref{Series})) we provide a short
discussion
on the theory of a higher-weight modules over Lie algebra.
We start with very well known Lie algebra ${\mathfrak g}{\mathfrak l}(n, {k})$.
\footnote{
The symbol $k$ denotes the field of real numbers $\mathbb R$
or the field $\mathbb C$ of complex numbers. In particular,
${\mathfrak g}{\mathfrak l}(n, {\mathbb C})$ is the Lie algebra of all complex $n\times n$
matrices with the operation $A, B \mapsto [A, B] = AB - BA$.
}
Results for ${\mathfrak g}{\mathfrak l}(n, {k})$
survive the passage to the limit $n\rightarrow \infty$, if one assumes that
${\mathfrak g}{\mathfrak l}(\infty, {k})$ is the Lie algebra of infinite finitary matrices,
which means $\bigcup_n {\mathfrak g}{\mathfrak l}(n, {k})$.
In this remark we deal with the Lie algebra ${\mathfrak g}{\mathfrak l}_{\widehat{\cJ}} (k)$
of generalized Jacobian matrices
\footnote{
The bilateral matrix $\Vert a_{ij}\Vert_{i,j\in {\mathbb Z}}$ is called a generalized Jacobian
matrix if it has a finite number of nonzero diagonals (that is, if there exists a
positive $N$ such that $a_{ij}=0$ for $\vert j-i\vert> N$). It is clear that the set
of generalized Jacobian matrices constitutes a Lie algebra, with respect to the usual
commutation rule.
}.
The algebra ${\mathfrak g}{\mathfrak l}_{\widehat{\cJ}} (k)$ can be considered as a nontrivial
one-dimensional central extension of the Lie algebra
${\mathfrak g}{\mathfrak l}_{\cJ}({k})$ (for details, see \cite{Fuks}). It is obvious that
${\mathfrak g}{\mathfrak l}_{\cJ}({k})\supset {\mathfrak g}{\mathfrak l}_{\cJ}(\infty, {k})$.
The importance of this Lie algebra stems from the following facts:
\begin{itemize}
\item{}
Many of the classical constructions of the theory of representations
of the Lie algebra ${\mathfrak g}{\mathfrak l}_{\cJ}({k})$ can be also applied to the
algebra ${\mathfrak g}{\mathfrak l}_{\widehat{\cJ}}({k})$. This creates a sizable supply of
${\mathfrak g}{\mathfrak l}_{\widehat{\cJ}}({k})$-modules.
\item{}

Important infinite-dimensional Lie algebras can be embedded in
${\mathfrak g}{\mathfrak l}_{\widehat{\cJ}}({k})$.
Thus, the already mentioned representations of
${\mathfrak g}{\mathfrak l}_{\widehat{\cJ}}({k})$  become representations of these algebras.
\item{}
The subalgebra of ${\mathfrak g}{\mathfrak l}_{\cJ} (k)$ composed of $n$-periodic matrices,
$\Vert a_{ij}\Vert$ with $a_{i+n, j+n}=a_{ij}$, is isomorphic to the algebra of currents
\cite{Fuks}
\footnote{
Recall that the space of smooth maps $\xX \rightarrow {\mathfrak g}$,
where $\xX$ is a smooth manifold and ${\mathfrak g}$ is a finite-dimensional Lie algebra,
with the $\cC^\infty$-topology and the commutator $[\mf, \mg](x) = [\mf(x), \mg(x)]$,
is the (topological) {\it current} Lie algebra and is denoted by ${\mathfrak g}^{\xX}$.
Together with the algebra ${\mathfrak g}^{S^1}$ \,($\xX=S^1$) one can consider its subalgebra
$({\mathfrak g}^{S^1})^{\rm pol}$, consisting of maps described by trigonometric polynomials.
For any commutative associative algebra $A$, the tensor product ${\mathfrak g}\otimes A$ is a
Lie algebra with respect to the commutators
$[\mg_1\otimes a_1, \mg_2\otimes a_2] = [\mg_1, \mg_2]\otimes a_1 a_2$; also
$({\mathfrak g}^{S^1})^{\rm pol} = {\mathfrak g}\otimes{\mathbb C}[t, t^{-1}]$.
}.
A non-trivial central extension of
${\mathfrak g}^X$ -- a Kac-Moody algebra -- is embedded in
${\mathfrak g}{\mathfrak l}_{\widehat{\cJ}} ({k})$. The Lie algebra
${L}^{\rm pol} = {\mathbb C}({\rm Vect}\,S^1)^{\rm pol}$ of complex polynomial vector fields
on the circle can be embedded in
${\mathfrak g}{\mathfrak l}_{\cJ}({k}= {\mathbb C})$. Recall that
${L}^{\rm pol}$ has a basis ${\bf e}_i$ and commutators of the form
\begin{equation}
[{\bf e}_i, {\bf e}_j]  =  (i-j) {\bf e}_{i+j}\,\,\,\,\,
(j\in {\mathbb Z}),
\,\,\,\,\,
{\bf e}_j  =   -x^{j+1}d/dx \,\,\,\,\,
{\rm on}\,\,\,\,\, {\mathbb C}\setminus \{0\}\,.
\label{basis}
\end{equation}
(The cohomologies of the algebra ${L}^{\rm pol}$ are known; in particular,
$H^2({L}^{\rm pol})= {\mathbb C}$.)
The Virasoro algebra is a Lie algebra over $\mathbb C$
with basis $L_n$ ($n\in {\mathbb Z}$), $c$. Because of
Eq.~(\ref{basis}), the Lie Virasoro algebra is a (universal)
central extension of the Lie algebra of holomorphic vector fields on
the punctured complex plane having finite Laurent series. It is for this reason that
the Virasoro algebra plays a key role in conformal field theory.
\end{itemize}

We briefly note  some elements of the representation theory of the
Virasoro algebra which are, in fact, very similar to those for
Kac-Moody algebras. Let us consider the highest representation of
the Virasoro algebra. Let $\MM(c, h)\, (c, h \in {\mathbb C})$ be
the Verma module over the Virasoro algebra.
The {\it conformal central charge} $c$ acts on $\MM(c,
h)$ as $cId$. As $[{\bf e}_0, {\bf e}_{-j}] = n {\bf e}_{-j}$,
${\bf e}_0$ is diagonalizable on $\MM(c, h)$, with spectrum $h+
{\mathbb Z}_{+}$ and eigenspace decomposition given by: $ \MM(c, h)
=\bigoplus_{j\in {\mathbb Z}_{+}} \MM(c, h)_{h+j}, $ where $\MM(c,
h)_{h+j}$ is spanned by elements of the basis $\{{\bf
e}_{-j_k}\}_{k=1}^n$ of $\MM(c, h)$. The number $ Z_j = {\rm dim}\,
\MM(c, h)_{h+j}, $ is the {\it classical partition function}. This
means that the Konstant partition function for the Virasoro
algebra is the classical partition function. On the other hand,
the partition functions can be rewritten in the form (Cf. Eq. (\ref{Series}))
\begin{equation}
{\rm Tr}_{\MM(c, h)}\, q^{{\bf e}_0} := \sum_{\lambda}{\rm dim}\,
\MM(c, h)_{\lambda}\,q^{\lambda} = q^h\prod_{j=1}^\infty (1-q^j)^{-1}\,.
\label{ch}
\end{equation}
The series ${\rm Tr}_{\mathcal V}\,q^{{\bf e}_0}$ is called the formal character of
the Virasoro-module ${\mathcal V}$. (A $\mathfrak g$-module ${\mathcal V} \in {C}$, where
$C$ is a category if: There is an expansion ${\mathcal V}=
\bigoplus_{\lambda \in {\mathfrak h}^*}{\mathcal V}_\lambda$
\, ($\mathfrak h$ is a Cartan subalgebra of $\mathfrak g$) and
$e_\alpha^{(i)}{\mathcal V}_\lambda \subset {\mathcal V}_{\lambda+\alpha}$, where
$e_\alpha^{(i)}$ are root vectors correspond to root $\alpha$;
${\rm dim}\,{\mathcal V}_\lambda < \infty$ for all $\lambda$;
$D(\lambda) := \{ \lambda \in
{\mathfrak h}^*\mid {\mathcal V}_\lambda \neq 0\}\subset \bigcup_{i=1}^sD(\lambda_{i})$
for some $\lambda_1, \ldots, \lambda_s \in {\mathfrak h}^*$.)

\subsection{The polynomial ring $\Lambda(X)$}
\label{Ring}

Our aim in this Section is to exploit the Hopf algebra of the ring $\Lambda(X)$
of symmetric functions of the independent variables $(x_1, x_2, \ldots)$,
finite or contably infinite in number, that constitute the alphabet $X$.
In our notations and basic statements we shall mainly follow the lines of
article \cite{Fauser10}.

Let ${\mathbb Z}[x_1,\ldots,x_n]$ be the polynomial ring, or the
ring of formal power series, in $n$ commuting variables $x_1,\ldots,x_n$. The symmetric group
$S_n$ acting on  $n$ letters acts on this ring by permuting the variables.
For $\pi\in S_n$ and $f \in {\mathbb Z}[x_1,\ldots,x_n]$ we have
$
\pi f(x_1,\ldots,x_n) = f(x_{\pi(1)},\ldots,x_{\pi(n)}).
$
We are interested in the subring of functions invariant under this
action, $\pi f = f$, that is to say the ring of symmetric polynomials in
$n$ variables:
$
\Lambda(x_1,\ldots,x_n) = {\mathbb Z}[x_1,\ldots,x_n]^{S_n}.
$
This ring may be graded by the degree of the polynomials, so that
$
\Lambda(X)=\oplus_n\ \Lambda^{(n)}(X)
$,
where $\Lambda^{(n)}(X)$ consists of homogenous symmetric polynomials  in $x_1,\ldots,x_n$
of total degree $n$.

In order to work with an arbitrary number of variables, following
Macdonald~\cite{Macdonald}, we define the ring of symmetric functions
$\Lambda = \lim_{n\rightarrow\infty}\Lambda(x_1,\ldots,x_n)$ in its stable limit  ($n\rightarrow
\infty$). There exist various bases of $\Lambda(X)$:

{\bf (i)} A $\mathbb Z$ basis of $\Lambda^{(n)}$ is provided by the
monomial symmetric functions $\{m_\lambda\}$, where $\lambda$ is any partitions of $n$.

{\bf (ii)} The other (integral and rational) bases for $\Lambda^{(n)}$ are indexed by the
partitions $\lambda$ of $n$. There are the complete,
elementary and power sum symmetric functions bases defined multiplicatively in terms of
corresponding one part functions by:
$h_{\lambda}=h_{\lambda_1}h_{\lambda_2}\cdots h_{\lambda_n}$,
$e_{\lambda}=e_{\lambda_1}e_{\lambda_2}\cdots e_{\lambda_n}$ and
$p_{\lambda}=p_{\lambda_1}p_{\lambda_2}\cdots p_{\lambda_n}$ where the one part functions
are defined for $\forall n \in {\mathbb Z}_+$ by
\begin{equation}
h_n(X) = \sum_{i_1\leq i_2\cdots\leq i_n}x_{i_1}x_{i_2}\cdots x_{i_n},\,\,\,\,\,\,\,\,
e_n(X) = \sum_{i_1<i_2\cdots<i_n}x_{i_1}x_{i_2}\cdots x_{i_n},\,\,\,\,\,\,\,\,
p_n(X) = \sum_{i}x_i^n,
\end{equation}
with the convention $h_0 = e_0 = p_0 =1,\, h_{-n} = e_{-n} = p_{-n} = 0$.
Three of these bases are multiplicative,
with $h_{\lambda}=h_{\lambda_1}h_{\lambda_2}\cdots h_{\lambda_n}$,
$e_{\lambda}=e_{\lambda_1}e_{\lambda_2}\cdots e_{\lambda_n}$ and
$p_{\lambda}=p_{\lambda_1}p_{\lambda_2}\cdots p_{\lambda_n}$.
The relationships between the various bases we just mention at this stage by
the transitions
\begin{equation}
p_\rho(X) = \sum_{\lambda\,\vdash n}\chi_\rho^\lambda s_\lambda(X)
\,\,\,\,\,\,\, {\rm and} \,\,\,\,\,\,\,
s_\lambda(X) = \sum_{\rho\,\vdash n}\ \fz_\rho^{-1}\ \chi^\lambda_\rho\ p_\rho(X)\,.
\label{Eq-p-s}
\end{equation}
For each partition $\lambda$, the Schur function is defined by
\begin{equation}
s_\lambda(X)\equiv s_\lambda(x_1,x_2, \ldots, x_n) = \frac{\sum_{\sigma\in S_n}{\rm sgn}
(\sigma)X^{\sigma(\lambda+\delta)}}{\prod_{i<j}(x_i-x_j)}\,,
\end{equation}
where $\delta = (n-1,n-2,\ldots,1,0)$. In fact both $h_n$ and $e_n$ are special Schur functions,
$h_n = s_{(n)},\, e_n = s_{(1^n)}$, and their generating functions are expressed in terms of
the power-sum $p_n$:
\begin{equation}
\sum_{n\geq 0}h_nz^n = \exp (\sum_{n=1}^\infty(p_n/n)z^n), \,\,\,\,\,\,\,\,\,\,
\sum_{n\geq 0}e_nz^n = \exp (-\sum_{n=1}^\infty(p_n/n)(-z)^n)\,.
\end{equation}
The Jacobi-Trudi formula \cite{Macdonald} express the Schur functions in terms of $h_n$ or
$e_n$: $s_\lambda = {\rm det}(h_{\lambda_i-i+j}) = {\rm det}(e_{\lambda^\prime-i+j})$,
where $\lambda^\prime$ is the conjugate of $\lambda$.
An involution $\omega: \Lambda\rightarrow \Lambda$ can be defined by $\omega(p_n) = (-1)^{n-1}p_n$. Then it follows that
$\omega(h_n) = e_n$. Also we have $\omega(s_\lambda) = s_{\lambda^\prime}$.
$\chi^\lambda_\rho$ is the character of the irreducible representation
of the symmetric groups $S_n$ specified by $\lambda$ in the conjugacy class
specified by $\rho$. These characters satisfy the orthogonality conditions
\begin{eqnarray}
\sum_{\rho\,\vdash n}\ \fz_\rho^{-1}\ \chi^\lambda_\rho\ \chi^\mu_\rho = \delta_{\lambda,\mu}
\,\,\,\,\,\,\, {\rm and} \,\,\,\,\,\,\,
\sum_{\lambda\,\vdash n}\ \fz_\rho^{-1} \chi^\lambda_\rho \chi^\lambda_\sigma\ =
\delta_{\rho,\sigma}\,.
\label{Eq-chi-orth}
\end{eqnarray}
The significance of the Schur function basis lies in the fact that with respect
to the usual Schur-Hall scalar product
$\langle \cdot \,|\, \cdot \rangle_{\Lambda(X)}$ on $\Lambda(X)$ we have
\begin{eqnarray}
\langle s_\mu(X) \,|\, s_\nu(X) \rangle_{\Lambda(X)}
= \delta_{\mu,\nu}\,\,\,\,\,\, {\rm and}\,\,\,\,\, {\rm therefore}
\,\,\,\,\,
\langle p_\rho(X) \,|\, p_\sigma(X) \rangle_{\Lambda(X)}
= \fz_\rho \delta_{\rho,\sigma}\,,
\label{Eq-scalar-prod-p}
\end{eqnarray}
where $\fz_\lambda = \prod_i i^{m_i}m_i!$ for $\lambda = (1^{m_1}, 2^{m_2}, \cdots)$.

{\bf Algebraic properties of $\Lambda(X)$.}
The ring, $\Lambda(X)$, of symmetric functions over $X$ has a Hopf algebra
structure, and two further algebraic and two coalgebraic operations. For
notation and basic properties we refer the reader to
\cite{fauser:jarvis:2003a,fauser:jarvis:king:wybourne:2006a} and
references therein.

-- We indicate outer products on $\Lambda$ either by $m$, or with infix notation
using juxtaposition.

-- Inner products are denoted either by $\textsf{m}$ or as
infix by $\ast$

-- Plethysms (compositions) are denoted by $\circ$ or by
means of square brackets $[\,\,]$; plethysm coproduct is denoted by $\triangledown$.

-- The corresponding coproduct maps are specified by $\Delta$ for the outer coproduct.

-- Notation $\delta$ we use for the inner coproduct.
\footnote{In Sweedler notation the action of these coproducts is distinguished by means of
different brackets, round, square and angular, around the Sweedler indices --
the so-called Brouder-Schmitt convention.}

The coproduct coefficients themselves are obtained
from the products by duality using the Schur-Hall scalar product and the
self-duality of $\Lambda(X)$. For example, for all $A,B\in\Lambda(X)$:
\begin{eqnarray}
m(A\otimes B) & = &  AB; \,\,\,\,\,\,\,\,\,\,\,\,\,\,\,\,\,\,\,
\Delta(A) = A_{(1)}\otimes A_{(2)};
\nonumber \\
\textsf{m}(A\otimes B) & = & A\ast B; \,\,\,\,\,\,\,\,\,\,\,\,\,\,\,\,
\delta(A) = A_{[1]}\otimes A_{[2]};
\nonumber \\
A\circ B & = & A[B]; \,\,\,\,\,\,\,\,\,\,\,\,\,\,\,\,\, \triangledown (A) =
A_{\langle 1\rangle}\otimes
A_{\langle 2\rangle}\,.
\nonumber
\end{eqnarray}
In terms of the Schur function basis $\{s_\lambda\}_{\lambda\,\vdash
n,n\in\mathbb{Z}_+}$ the product and coproduct maps give
rise to the particular sets of coefficients specified as follows:
\begin{eqnarray}
s_\mu s_\nu & = & \sum_\lambda c^\lambda_{\mu,\nu}s_\lambda \,;
\,\,\,\,\,\,\,\,\,
\Delta(s_\lambda)
= s_{\lambda_{( 1 ) }}\otimes s_{\lambda_{( 2 )}}
=\sum_{\mu,\nu}c^\lambda_{\mu,\nu} s_\mu\otimes s_\nu \,;
\nonumber \\
s_\mu\ast s_\nu & = & \sum_\lambda g^\lambda_{\mu,\nu}s_\lambda \,;
\,\,\,\,\,\,\,\,\,
\delta(s_\lambda) = s_{\lambda_{[1]}}\otimes s_{\lambda_{[2]}}\,\,\,\,\,
=\sum_{\mu,\nu} g^\lambda_{\mu,\nu}s_\mu\otimes s_\nu\,;
\nonumber \\
s_\mu[s_\nu] & = & \sum_\lambda p_{\mu, \nu}^\lambda s_\lambda \,;
\,\,\,\,\,\,\,\,\,
\triangledown (s_\lambda) \,\, =  s_{\lambda_{\langle 1\rangle}}\otimes
s_{\lambda_{\langle 2\rangle}} \, = \sum_{\mu,\nu}  p_{\mu, \nu}^\lambda s_\mu\otimes s_\nu\,.
\end{eqnarray}
Here the $c^\lambda_{\mu,\nu}$ are Littlewood-Richardson coefficients,
the $g^\lambda_{\mu,\nu}$ are Kronecker coefficients and the
$p^\lambda_{\mu,\nu}$ are plethysm coefficients. All these coefficients
are non-negative integers. The Littlewood-Richardson coefficients
can be obtained, for example, by means of the Littlewood-Richardson
rule~\cite{littlewood:richardson:1934a,littlewood:1950a} or the hive
model~\cite{buch:2000a}. The Kronecker coefficients may determined directly
from characters of the symmetric group or by exploiting the Jacobi-Trudi
identity and the Littlewood-Richardson rule, while
plethysm coefficients have been the subject of a variety methods of
calculation~\cite{littlewood:1950b,chen:garsia:remmel:1984a}.
Note that the above sums are finite, since
$c^\lambda_{\mu,\nu} \geq 0$ iff $\vert\lambda\vert=\vert\mu\vert +\vert\nu\vert$;
$g^\lambda_{\mu,\nu} \geq 0$ iff $\vert\lambda\vert=\vert\mu\vert =\vert\nu\vert$;
$p^\lambda_{\mu,\nu} \geq 0$ iff $\vert\lambda\vert=\vert\mu\vert\,\vert\nu\vert$.

The Schur-Hall scalar product may be used to define skew Schur functions
$s_{\lambda/\mu}$ through the identities
$
c^\lambda_{\mu,\nu}
= \langle s_\mu\, s_\nu \vert s_\lambda \rangle  = \langle s_\nu\,\vert s_\mu^\perp (s_\lambda)
\rangle = \langle s_\nu \vert s_{\lambda/\mu} \rangle\,,
$
so that
$
s_{\lambda/\mu} =  \sum_\nu\ c^\lambda_{\mu,\nu}\ s_\nu\,.
$
Within the outer product Hopf algebra we have a unit $\rm Id$,
a counit $\varepsilon$ and an antipode \textsf{S} such that
\footnote{
Macdonald uses the involution $\omega$ which differs from the antipode
by a sign factor: $\textsf{S}(s_\lambda)=(-1)^{\ell(\lambda)}\omega(s_\lambda)$.
It is, however, convenient to employ the antipode if Hopf algebra structures
are in use.}: ${\rm Id}(1) = s_0$;\, $\varepsilon(s_\lambda) =\delta_{\lambda,(0)}$\,;
$\textsf{S}(s_\lambda) = (-1)^{|\lambda|} s_{\lambda'}\,$.

In what follows we shall make considerable use of several infinite series
of Schur functions. The most important of these are the mutually inverse
pair defined by
\begin{eqnarray}
 M(t;X) & = & \prod_{i\geq 1} (1-t\,x_i)^{-1} = \sum_{m\geq 0} h_m(X)t^m\,
\label{Eq-M},\\
L(t;X) & = & \prod_{i\geq 1} (1-t\,x_i)\,\,\,\,\,\,
= \, \sum_{m\geq 0}(-1)^m e_m(X) t^m\,,
\label{Eq-L}
\end{eqnarray}
where as Schur functions $h_m(X)=s_{(m)}(X)$ and $e_m(X)=s_{(1^m)}(X)$.
\footnote{It might be noted that in Macdonald's notation and $\lambda$-ring
notation $M(t;X)=H(t)=\sigma_t(X)$ and $L(t;X)=E(-t)=\lambda_{-t}(X)$.}
For convenience, in the case $t=1$ we write $M(1;X)=M(X)$ and $L(1;X)=L(X)$.

\subsection{Plethysms}
\label{Plethysm}

Plethysms are defined as compositions whereby for any $A,B\in\Lambda(X)$; the
plethysm $A[B]$ is $A$ evaluated over an alphabet $Y$ whose letters are the
monomials of $B(X)$, with each letter repeated as many times as the
multiplicity of the corresponding monomial. For example, the Schur function plethysm
is defined by $s_\lambda[s_\mu](X) = s_\lambda(Y)$, where $Y=s_\mu(X)$.

For all $A,B,C\in \Lambda(X)$ we have the following rules, due to
Littlewood~\cite{littlewood:1950a}, for manipulating plethysms:
\footnote{
To these we can add, see~\cite{fauser:jarvis:king:wybourne:2006a}:
$
A[-B]  =  (\textsf{S}(A))[B];\, A[\textsf{S}(B)] = {\textsf{S}}(A[B]);\,
A[\Delta(B)]  =  \Delta(A[B]);\, A[\delta(B)] = \delta(A[B])$,\,
and the plethysm of a tensor product:
$
A[B\otimes C]=A_{[1]}[B]\otimes A_{[2]}[C].
$
}
\begin{eqnarray}
\label{littlewood:plethysm}
(A+B)[C] & = & A[C] + B[C]\,; \,\,\,\,\,\,\,\,\,\, A[B + C] = A_{(1)}[B] A_{(2)}[C]\,;
\nonumber \\
(AB)[C]  & = & A[C]B[C]\,; \,\,\,\,\,\,\,\,\,\,\,\,\,\,\,\,\,\,\,\,\,\,\,\,\,
A[BC] = A_{[1]}[B]A_{[2]}[C]\,;
\nonumber \\
 A[B[C]]  & = & (A[B])[C]\,.
\end{eqnarray}
These rules enable us to evaluate plethysms not only of outer and inner
products but also of outer and inner coproducts.

{\bf The Cauchy kernel.}
It is often convenient to represent an alphabet in an additive manner
$X$, as itself an element of the ring $\Lambda(X)$ in the
sense that
$x_1+x_2+\cdots\ = h_1(X) = e_1(X) = p_1(X) = s_{(1)}(X).
$
As elements of $\Lambda(X)\otimes\Lambda(Y)$ we have
$X + Y = \sum_{j=1}x_j+\sum_{j=1}y_j$,\, $XY = \sum_jx_j\sum_jy_j$.
With this notation, the outer coproduct gives
\begin{eqnarray}
\Delta(M) & = & M_{(1)}\otimes M_{(2)} = M\otimes M;
\,\,\,\,\,\,\,\,\,\,
M(X\!\!+\!\!Y) = \prod_i (1-x_i)^{-1}\,\prod_j\, (1-y_j)^{-1}\,;
\nonumber \\
\Delta(L) & = & L_{(1)}\otimes L_{(2)} = L\otimes L;
\,\,\,\,\,\,\,\,\,\,\,\,\,\,\,\,\,\,\,\,\,
L(X\!\!+\!\!Y) = \prod_i (1-x_i)\, \prod_j\,(1-y_j) \,,
\nonumber
\end{eqnarray}
so that $M(X\!\!+\!\!Y)=M(X)\,M(Y)$ and $L(X\!\!+\!\!Y)=L(X)\,L(Y)$.
For the inner coproduct:
\begin{eqnarray}
  \delta(M)
     &=M_{[1]}\otimes M_{[2]};
\,\,\,\,\,\,\,\,\,\,
       M(XY)
     &= \prod_{i,j} (1-x_iy_j)^{-1}\,;
   \nonumber \\
  \delta(L)
     &=L_{[1]}\otimes L_{[2]};
\,\,\,\,\,\,\,\,\,\,
       L(XY)
     &= \prod_{i,j} (1-x_iy_j) \,.
     \nonumber
\end{eqnarray}
The expansions of the products on the right hand sides of these expressions is
effected remarkably easily by evaluating the inner coproducts on the left:
\begin{eqnarray}
\delta(M) & = & \sum_{k\geq 0}\, \delta(h_k)
= \sum_{k\geq 0} \sum_{\lambda\vdash k}\ s_\lambda \otimes s_\lambda\,;
  \nonumber \\
\delta(L) & = & \sum_{k\geq0}\,(-1)^k\,\delta(e_k)
= \sum_{k\geq0}\, (-1)^k\, \sum_{\lambda\vdash k}\
s_\lambda\otimes s_{\lambda^\prime}\,.
  \nonumber
\end{eqnarray}
This gives immediately the well known Cauchy and Cauchy-Binet formulas:
\begin{eqnarray}
M(XY) & = & \prod_{i,j}\ (1-x_i\,y_j)^{-1}
= \sum_\lambda\ s_\lambda(X)\, s_\lambda(Y)\,;
\label{Eq-Cauchy}
    \\
L(XY) & = & \prod_{i,j}\ (1-x_i\,y_j)
= \sum_\lambda\ (-1)^{|\lambda|} s_\lambda(X)\,s_{\lambda^\prime}(Y) \,.
\label{Eq-dual-Cauchy}
\end{eqnarray}
The Cauchy kernel, $M(XY)$, is a dual version of the Schur-Hall scalar
product; indeed:
\begin{eqnarray}
s_\mu(X)\ M(XY) & = & \sum_\lambda\, \sum_\nu\, c^\nu_{\mu,\lambda}\ s_\nu(X)\, s_\lambda(Y)
  \nonumber \\
& = &
\sum_\nu\, s_\nu(X)\, s_{\nu/\mu}(Y) = s^\perp_\mu(Y)\,(M(XY)).
\label{Eq-sM-sperpM}
\end{eqnarray}
Generally speaking, for any $F(X)\in \Lambda(X)$ with dual $F^\perp(X)$, by
linearly extending the above result we have $F(X)\ M(XY) = F^\perp(Y)\,(M(XY))$.

\subsection{Multipartite generating functions and symmetric functions}
\label{Mult}

We note here an alternative definition of plethisms, involving a particular substitution process.
For  $M(z;X) =\prod_{i\geq1} (1-z\,x_i)^{-1}$, \, $L(z;X)  =  \prod_{i\geq1} (1-z\,x_i)
= M^{-1}(z;X)$ it follows that
\begin{eqnarray}
&&
{\rm log}\, M(z;X)  =  -\sum_{i\geq1} \ {\rm log}\, (1-z\,x_i)
 = \sum_{i\geq1}\sum_{k\geq 1}\frac{(zx_i)^k}{k}
= \sum_{k\geq1} \ \frac{z^k}{k}\, p_k(X)\,,
\label{log1}
\\
&&
{\rm log}\, L(z;X)  =  - {\rm log}\, M(x;Z) = - \sum_{k\geq1} \ \frac{z^k}{k}\, p_k(X)\,,
\label{log2}
\\
&&
M(z;X)  =  \exp \left( \sum_{k\geq1} \ (z^k/k)\, p_k(X) \right),
L(z;X) = \exp \left( - \sum_{k\geq1} \ (z^k/k)\, p_k(X) \right).
\end{eqnarray}
Let for all positive integers $k$,\,
$
p^\perp_k(X) = k \frac{\partial}{\partial p_k(X)},
$
then \,
$
L^\perp(z;X) = \exp( -\sum_{k\geq 1} z^k\frac{\partial}{\partial p_k(X)}).
$

Given a symmetric function $F(X) = \sum_{\alpha}C_\alpha X^\alpha$
expressed in terms of monomials $X^\alpha = x_1^{\alpha_1}x_2^{\alpha_2} \cdots$
define new variables $\{y_1, y_2, \cdots\}$ by
\begin{equation}
\prod_i (1+ zy_i) = \prod_\alpha(1+ zx^\alpha )^{C_\alpha}.
\label{s1}
\end{equation}
This allows one to define the plethysm of two symmetric functions $F, G \in \Lambda(X)$
by $(F\otimes G)(X) = G(y_1, y_2, \cdots).$
This operation is right-distributive, but not left-distributive \cite{Baker,Littlewood}:
\begin{equation}
(F + G)\otimes s_\lambda = \sum_\rho (F\otimes s_{\lambda/\rho})(G\otimes s_\rho)\,.
\end{equation}
\footnote{
An important result in the theory of plethysms is:
$
s_\lambda\otimes s_\mu = \sum_\rho a_{\lambda\mu}^\rho s_\rho,
$
where $a_{\lambda\mu}^\rho$ are non-negative integers, and the sum is over all partitions
of weight $|\rho|= |\lambda|\cdot |\mu|$.
There is a standard method for recursively computing plethysms based on the
identity \cite{Littlewood}:
$
\sum_{n=0}^{|\lambda||\mu|}D_{(n)}(s_\lambda\otimes s_\mu) = (\sum_{m=0}^{|\mu|}s_{\mu/(m)})
\otimes s_{\lambda}\,.
\label{s}
$
Here $D_\lambda$ is the adjoint (skew) operator $D_\lambda s_\sigma = s_{\sigma/\lambda}$.
}
By taking the logarithm of (\ref{s1}) as it has been done in (\ref{log1}), (\ref{log2}),
it follows that $p_n(Y) = F(X^n)$.
In order to calculate the plethysm $s_\lambda\otimes s_\mu$, one can expresses
$s_\mu(X)$ as a multinomial in the power
sums $p_1(x), p_2(x), \cdots$, and then makes the substitution $p_j(X) \rightarrow s_\lambda(x^j)$.
Thus a knowledge of how to express $s_\lambda(x^j)$ in terms on $S$-functions with argument $X$,
along with the Littlewood{ Richardson rule for multiplying the $S$-functions together,
is suficient to be able to calculate any plethysm.

Let us consider, for any ordered $r$-tuple of nonnegative integers not all zeros,
$(k_1, k_2, \ldots ,k_r)={\overrightarrow{k}}$
(referred to as "$r$-partite" or {\it multipartite} numbers), the
(multi)partitions, i.e. distinct representations of $(k_1, k_2, \ldots ,k_r)$ as sums
of multipartite numbers. Let us call
${\mathcal C}_-^{(\cO,r)}({\overrightarrow{k}}) = {\mathcal C}_-^{r}(\cO;k_1, k_2 , \cdots , k_r)$
the number of such multipartitions, and introduce in addition the symbol
${\mathcal C}_+^{(\cO,r)} ({\overrightarrow{k}})=
{\mathcal C}_+^{(r)}(\cO;k_1, k_2 , \cdots , k_r)$.
Their generating functions are defined by
\begin{eqnarray}
{\mathcal F}({\cO;X}) &: = & \prod_{{\overrightarrow{k}}\geq 0} \left( 1- \cO x_1^{k_1}x_2^{k_2}
\cdots x_r^{k_r}\right)^{-1}
= \sum_{{\overrightarrow{k}}\geq 0}{\mathcal C}_-^{(\cO,r)}({\overrightarrow{k}})
x_1^{k_1}x_2^{k_2}\cdots x_r^{k_r}\,,
\label{PF1}
\\
{\mathcal G }(\cO;X) &: = & \prod_{{\overrightarrow{k}}\geq 0} \left( 1 + \cO x_1^{k_1}x_2^{k_2}
\cdots x_r^{k_r}\right)
= \sum_{{\overrightarrow{k}}\geq 0}{\mathcal C}_+^{(\cO,r)}({\overrightarrow{k}})
x_1^{k_1}x_2^{k_2}\cdots x_r^{n_r}\,.
\label{PF2}
\end{eqnarray}
Therefore,
\begin{eqnarray}
{\rm log}\, {\mathcal F}(\cO;X) & = & - \sum_{{\overrightarrow{k}}\geq 0} {\rm log}
\left(1- \cO x_1^{k_1}x_2^{k_2}\cdots x_r^{k_r}\right)
=  \sum_{{\overrightarrow{k}}\geq 0} \sum_{m=1}^\infty \frac{\cO^m}{m}
x_1^{mk_1}x_2^{mk_2}\cdots x_r^{mk_r}
\nonumber \\
& = &
\sum_{m =1}^\infty \frac{\cO^m}{m}
(1-x_1^m)^{-1}(1-x_2^m)^{-1} \cdots (1-x_r^m)^{-1}
\nonumber \\
& = &
\sum_{m=1}^\infty \frac{\cO^m}{m}
\prod_{j= 1}^r (1-x_j^m)^{-1},
\\
{\rm log}\,{\mathcal G}(-\cO;X) & = & {\rm log}\,{\mathcal F}(\cO;X)\,.
\end{eqnarray}
Finally,
\begin{eqnarray}
\!\!\!\!\!\!\!\!\!\!\!\!\!
{\mathcal F}(\cO;X) & = &
\sum_{{\overrightarrow{k}}\geq 0}{\mathcal C}_-^{(\cO,r)}({\overrightarrow{k}})
x_1^{k_1}x_2^{k_2}\cdots x_r^{k_r}
= \exp\left( \sum_{m=1}^\infty \frac{\cO^m}{m}
\prod_{j=1}^r (1-x_j^m)^{-1}\right),
\\
\!\!\!\!\!\!\!\!\!\!\!\!\!
{\mathcal G }(\cO;X) & = &
\sum_{{\overrightarrow{k}}\geq 0}{\mathcal C}_+^{(\cO,r)}({\overrightarrow{k}})
x_1^{k_1}x_2^{k_2}\cdots x_r^{n_r}
= \exp\left( \sum_{m=1}^\infty \frac{(-\cO)^m}{m}
\prod_{j=1}^r (1-x_j^m)^{-1}\right).
\end{eqnarray}
\\

{\bf Restricted specialization.}
Setting $X= (x_1, x_2, \ldots,x_r, 0, 0, \ldots)$; for finite additive manner let
$p_m(X):= \prod_{j=1}^r(1-x_j^m)^{-1}$, then
\begin{eqnarray}
\!\!\!\!\!\!\!\!\!\!\!\!\!
{\mathcal F}(\cO;X) & = &
\sum_{{\overrightarrow{k}}\geq 0}{\mathcal C}_-^{(\cO,r)}({\overrightarrow{k}})
x_1^{k_1}x_2^{k_2}\cdots x_r^{k_r}
= \exp\left( \sum_{m=1}^\infty \frac{\cO^m}{m}
\exp \left(\sum_{k\geq 1}^r\frac{1}{k}p_m(X))\right)\right).
\\
\!\!\!\!\!\!\!\!\!\!\!\!\!
{\mathcal G }(\cO;X) & = &
\sum_{{\overrightarrow{k}}\geq 0}{\mathcal C}_+^{(\cO,r)}({\overrightarrow{k}})
x_1^{k_1}x_2^{k_2}\cdots x_r^{n_r}
= \exp\left( \sum_{m=1}^\infty \frac{(-\cO)^m}{m}
\exp \left(\sum_{k\geq 1}^r\frac{1}{k}p_m(X))\right)\right).
\end{eqnarray}

It is known that the Bell polynomials are very useful in many problems in combinatorics.
We would like to note their
application in multipartite partition problem \cite{Andrews}.
The Bell polynomials technique can be used for the
calculation ${\mathcal C}_-^{(r)}({\overrightarrow{k}})$ and
${\mathcal C}_+^{(r)}({\overrightarrow{k}})$.
Let
\begin{eqnarray}
&& {\mathcal F}(\cO;X) := 1 + \sum_{j=1}^\infty {\mathcal P}_j(x_1,x_2, \ldots, x_r)\cO^j,
\,\,\,\,\,\,\,\,\,\,
{\mathcal P}_j  =  1+ \sum_{{\overrightarrow{k}}> 0}P({\overrightarrow{k}}; j)
x_1^{n_1}\cdots x_r^{n_r},
\label{Fu}
\\
&& {\mathcal G}(\cO;X)  :=  1 + \sum_{j=1}^\infty {\mathcal Q}_j(x_1,x_2, \ldots, x_r)\cO^j,
\,\,\,\,\,\,\,\,\,\,
{\mathcal Q}_j  =  1+ \sum_{{\overrightarrow{k}}> 0}Q({\overrightarrow{k}}; j)
x_1^{n_1}\cdots x_r^{n_r}.
\label{Gu}
\end{eqnarray}
Useful expressions for the recurrence relation of the Bell polynomial
$Y_{n}(g_1, g_2, \ldots , g_{n})$
and generating function ${\mathcal B}(\cO)$ have the forms \cite{Andrews}:
\begin{eqnarray}
&& Y_{n+1}(g_1, g_2, \ldots , g_{n+1}) = \sum_{k=0}^n  \begin{pmatrix} n\cr k\end{pmatrix}
Y_{n-k}(g_1, g_2, \ldots , g_{n-k})g_{k+1},
\label{B2}
\end{eqnarray}
${\mathcal B}(\cO) = \sum_{n=0}^\infty Y_n \cO^n/n! \Longrightarrow
{\rm log}\,{\mathcal B}(\cO)= \sum_{n=1}^\infty g_n \cO^n/n!.
$
To verify the last formula we need to differentiate with respect to $\cO$ and observe
that a comparison of the coefficients of ${\cO}^n$ in the resulting equation produces an
identity equivalent to (\ref{B2}). From Eq. (\ref{B2}) one can obtain the following explicit
formula for the Bell polynomials (it is known as Faa di Bruno's formula)
\begin{equation}
Y_{n}(g_1, g_2, \ldots , g_{n}) = \sum_{{\bf k}\,\vdash\, n}\frac{n!}{k_1!\cdots k_n!}
\prod_{j=1}^n\left(\frac{g_j}{j!}\right)^{k_j}\!\!.
\end{equation}
The following result holds (see for detail \cite{Andrews}):
\begin{eqnarray}
{\mathcal P}_j  & = & \frac{1}{j!}Y_j \left( 0!p_1(X),\,\, 1!p_2(X)\,\,,
\ldots , \,\,(j-1)!p_j(X)\right),
\label{P1}
\\
{\mathcal Q}_j  & = & \frac{1}{(-1)^jj!}Y_j \left( -0!p_1(X),\,\,
-1!p_2(X)\,\,, \ldots , \,\,-(j-1)!p_j(X)\right).
\label{Q1}
\end{eqnarray}

For some specializations, when $X = q^\rho = (q, q^2, \ldots, q^r)$ we get
\begin{eqnarray}
{\mathcal F}(\cO;X) & = &
\prod_{{\overrightarrow{k}}\geq 0} \left( 1- \cO q^{k_1+k_2+\cdots + k_r}\right)^{-1}
= \exp\left( - \sum_{m=1}^\infty \frac{\cO^m}{m}\prod_{\ell = 1}^r(1-q^{\ell m})^{-1}\right),
\\
{\mathcal G}(\cO;X) & = &
\prod_{{\overrightarrow{k}}\geq 0} \left( 1+ \cO q^{k_1+k_2+\cdots + k_r}\right)
= \exp\left( - \sum_{m =1}^\infty \frac{(-\cO)^m}{m}\prod_{\ell = 1}^r(1-q^{\ell m})^{-1}\right).
\end{eqnarray}

{\bf Spectral functions of hyperbolic geometry.}
Let us begin by explaining the general lore on the characteristic classes and $\mg$-structure
on compact groups.
\begin{Statement}
Suppose that $\mathfrak g$ is the Lie algebra of a Lie group $G$. Let us
consider the pair $({H}, G)$ of Lie groups, where $H$ is a closed
subgroup of $G$ with normalizer subgroup $N_H\subset G$. Then the
pair $({H}, G)$ with the discrete quotient group $N_H/{ H}$ corresponds to the
inclusion ${\mathfrak
g}\hookrightarrow W_n$, where $W_n$ is the Lie algebra of formal vector
fields in $n = {\rm dim} \,G/H$ variables, while the homogeneous space $G/H$
possesses a canonical $\mathfrak g$-{structure} $\omega$. Combining this $\mg$-structure with the
inclusion $\mg\hookrightarrow W_n$, one obtains a $W_n$-structure on
the quotient space $G/\Gamma$ for any discrete subgroup $\Gamma$ of the Lie group $G$; this is
precisely the $W_n$-structure which corresponds to the $H$-equivariant foliation
of $G$ by left cosets of $\Gamma$~{\rm \cite{Fuks}}.
The homomorphism
$
{{\rm char}_\omega :} \, H^\sharp (W_n)\rightarrow H^\sharp (G/\Gamma,
{\mathbb R})
$
associated with characteristic classes of $W_n$-structures
decomposes into the composition of two homomorphisms
$
H^\sharp (W_n)\rightarrow H^\sharp ({\mathfrak g})
$ and
$
H^\sharp ({\mathfrak g}) \rightarrow H^\sharp (G/\Gamma, {\mathbb R})$;
the first homomorphism is independent of $\Gamma$
and is induced by the inclusion ${\mathfrak g}\hookrightarrow W_n$,
while the second homomorphism is independent of $H$ and corresponds to the
canonical homomorphism which determines
the characteristic classes of the canonical $\mathfrak g$-structure $\omega$ on $G/\Gamma$.
If the group $G$ is semi-simple, then the Lie algebra $\mathfrak g$ is
unitary and $G$ contains a discrete subgroup $\Gamma$ for which
$G/\Gamma$ is compact; for appropriate choice of $\Gamma$ the kernel of the homomorphism
$H^\sharp (W_n)\rightarrow H^\sharp (G/\Gamma, {\mathbb R})$
coincides with the kernel of the homomorphism
$H^\sharp (W_n)\rightarrow H^\sharp ({\mathfrak g})$.

In our applications we shall  consider a compact hyperbolic three-manifold $G/\Gamma$
with $G = SL(2, {\mathbb C})$.
By combining the characteristic class representatives of field theory elliptic
genera with the homomorphism
$
{{\rm char}_\omega}
$,
one can compute quantum partition functions in terms of the
spectral functions of hyperbolic three-geometry {\rm\cite{BCST}}.
\end{Statement}

Let us introduce next the Ruelle spectral function ${\mathcal R}(s)$ associated with
hyperbolic three-geometry \cite{BB,BCST}. The function ${\mathcal R}(s)$ is an
alternating product of more complicate factors, each of which is so-called Patterson-Selberg
zeta-functions $Z_{\Gamma^\gamma}$ (see Sect. \ref{Symmetry} and \cite{Patterson}).
Functions ${\mathcal R}(s)$ can be continued meromorphically to
the entire complex plane $\mathbb C$
\begin{eqnarray}
\prod_{n=\ell}^{\infty}(1- q^{an+\varepsilon})
& = & \prod_{p=0, 1}Z_{\Gamma^\gamma}(\underbrace{(a\ell+\varepsilon)(1-i\varrho(\vartheta))
+ 1 -a}_s + a(1 + i\varrho(\vartheta)p)^{(-1)^p}
\nonumber \\
& = &
\cR(s = (a\ell + \varepsilon)(1-i\varrho(\vartheta)) + 1-a),
\label{R1}
\\
\prod_{n=\ell}^{\infty}(1+ q^{an+\varepsilon})
& = &
\prod_{p=0, 1}Z_{\Gamma^\gamma}(\underbrace{(a\ell+\varepsilon)(1-i\varrho(\vartheta)) + 1-a +
i\sigma(\vartheta)}_s
+ a(1+ i\varrho(\vartheta)p)^{(-1)^p}
\nonumber \\
& = &
\cR(s = (a\ell + \varepsilon)(1-i\varrho(\vartheta)) + 1-a + i\sigma(\vartheta))\,,
\label{R2}
\end{eqnarray}
\begin{eqnarray}
\prod_{n=\ell}^{\infty}(1-q^{an+ \varepsilon})^{bn} & = &
\cR(s=(a\ell + \varepsilon)(1-i\varrho(\vartheta))+1-a)^{b\ell}
\nonumber \\
& \times &
\!\!\!
\prod_{n=\ell+1}^{\infty}
\cR(s=(an + \varepsilon)(1-i\varrho(\vartheta))+1-a)^{b}\,,
\label{RU1}
\\
\prod_{n=\ell}^{\infty}(1+q^{an+ \varepsilon})^{bn} & = &
\cR(s=(a\ell + \varepsilon)(1-i\varrho(\vartheta))+1-a+ i\sigma(\vartheta))^{b\ell}
\nonumber \\
& \times &
\!\!\!
\prod_{n=\ell+1}^{\infty}
\cR(s=(an + \varepsilon)(1-i\varrho(\vartheta))+1-a+ i\sigma(\vartheta))^{b}\,,
\label{RU2}
\end{eqnarray}
being $q\equiv e^{2\pi i\vartheta}$, $\varrho(\vartheta) =
{\rm Re}\,\vartheta/{\rm Im}\,\vartheta$,
$\sigma(\vartheta) = (2\,{\rm Im}\,\vartheta)^{-1}$,
$a$ is a real number, $\varepsilon, b\in {\mathbb C}$, $\ell \in {\mathbb Z}_+$.
\\

Obviously, \, $\prod_{\ell = 1}^r(1-q^{\ell m}) \equiv
\prod_{\ell = 1}^\infty(1-q^{\ell m})\prod_{\ell = r+1}^\infty(1-q^{\ell m})^{-1}$ and
\begin{eqnarray}
{\mathcal F}(\cO;X) & = &
\prod_{{\overrightarrow{k}}\geq 0} \left( 1- \cO q^{k_1+k_2+\cdots + k_r}\right)^{-1}
\nonumber \\
& = &
\exp\left( - \sum_{m=1}^\infty \frac{\cO^m}{m}
\left(\frac{\cR(s= -im\varrho(\vartheta)(r+1)+mr+1)}
{\cR(s = -im\varrho(\vartheta)+ 1)}\right)\right),
\label{F1}
\\
{\mathcal G}(\cO;X) & = &
\prod_{{\overrightarrow{k}}\geq 0} \left( 1+ \cO q^{k_1+k_2+\cdots + k_r}\right)
\nonumber \\
& = &
\exp\left( - \sum_{m =1}^\infty \frac{(-\cO)^m}{m}
\left(\frac{\cR(s= -im\varrho(\vartheta)(r+1)+mr+1)}
{\cR(s = -im\varrho(\vartheta)+ 1)}\right)\right)\,.
\label{G1}
\end{eqnarray}
\\

{\bf Hierarchy.}
Setting $\cO q^{k_0+k_{1}+\ldots + k_{r}}= \cO_{{\overrightarrow{k}}}q^{k_{0}}$
with $\cO_{{\overrightarrow{k}}}=
\cO q^{k_{1}+\ldots +k_{r}}$\, (${\overrightarrow{k}} =
\left( k_{1},\ldots ,k_{r}\right))$ we get
\begin{equation}
Z_{2}\left( \cO_{{\overrightarrow{k}}},q\right) =\prod_{k_{0}=0}^{\infty}
\left[1-\cO_{\overrightarrow{k}}q^{k_{0}}\right]^{-1} =
[(1-\cO_{\overrightarrow{k}}){\mathcal R}(s= (k_{1}+\ldots + k_{r})(1-i\varrho(\tau)))]^{-1}\,.
\label{CFT_2}
\end{equation}
Therefore the infinite products
$\prod_{k_r=0}^\infty\prod_{k_{r-1}=0}^\infty\cdots\prod_{k_1=0}^\infty\prod_{k_0=0}^\infty
(1-q^{k_0+k_1+\cdots+k_r})^{-1}$ can be factorized as $
\prod_{{\overrightarrow{k}}\geq {\overrightarrow{0}}} Z_{2}\left(
\cO_{{\overrightarrow{k}}},q\right). $ We can treat this
factorization as a product of $r$ copies, each of them is
$Z_{2}\left( \cO_{{\overrightarrow{k}}},q\right)$ and corresponds
to a free two-dimensional conformal field theory.

\subsection{Characters, branching rules and vertex operator traces}
\label{Character}

Let us discuss Schur functions interpretation as universal characters \cite{King}.
In this section our aim is try to make clear the connection between the Hopf algebraic
approach and the group theory. Our basic starting point is very well known Weyl's character
formula
\begin{equation}
\ch(\Lambda) = \frac{\sum_{w\in W} \varepsilon(w)e^{w(\Lambda+\rho)}}
{\sum_{w\in W} \varepsilon(w)e^{w(\rho)}}\,.
\end{equation}
Here $\Lambda$ is the highest weight vector, $\rho$ is half the sum of
the positive roots and $W$ is the appropriate Weyl group with
$\varepsilon$ the sign of $w$.

{\bf Characters of the classical groups.} Recall that the Cartan classification of the
simple complex classical
Lie groups is given by the series $A_n$, $B_n$, $C_n$ and $D_n$ (not to be confused with Schur
function series). These series correspond to the complexified versions of the
groups $SU(n+1)$, $SO(2n+1)$, $Sp(2n)$ and $SO(2n)$, which can be
considered as subgroups of unitary groups $U(N)$ for
$N=n+1,2n+1,2n$ and $2n$. Let us denote eigenvalues as $x_k = \exp(i\varphi_k)$; one can
write the eigenvalues of group elements as follows:
\begin{eqnarray}
GL(n):  && \,\,\,\,\, x_1,x_2,\ldots, x_n \,\,\,\,\,\,\, {\rm and}
\,\,\,\,\,\,\, x_1x_2\cdots x_n \neq 0.
\nonumber \\
SL(n):  && \,\,\,\,\, x_1,x_2,\ldots, x_n \,\,\,\,\,\,\, {\rm and}
\,\,\,\,\,\,\, x_1x_2\ldots x_n=1.
\nonumber \\
SU(n+1): && \,\,\,\,\, x_1,x_2,\ldots, x_n \,\,\,\,\,\,\, {\rm and}
\,\,\,\,\,\,\, x_1x_2\cdots x_n=1.
\nonumber \\
O(2n+1)\backslash SO(2n+1):
&& \,\,\,\,\, x_1,x_2,\ldots,x_n,\, {x}_1^{-1}, {x}_2^{-1},\ldots , {x}_n^{-1},-1.
\nonumber \\
O(2n)\backslash SO(2n):
&& \,\,\,\,\, x_1,x_2,\ldots,x_n,\, {x}_1^{-1}, {x}_2^{-1},\ldots , {x}_{n-1}^{-1},-1.
\nonumber \\
SO(2n):
&&  \,\,\,\,\, x_1,x_2,\ldots,x_n,\,{x}_1^{-1}, {x}_2^{-1},\ldots, {x}_n^{-1}.
\nonumber \\
SO(2n+1):
&& \,\,\,\,\, x_1,x_2,\ldots,x_n,\, {x}_1^{-1}, {x}_2^{-1},\ldots , {x}_n^{-1},1.
\nonumber \\
Sp(2n): &&  \,\,\,\,\, x_1,x_2,\ldots,x_n,\,{x}_1^{-1}, {x}_2^{-1},\ldots, {x}_n^{-1}.
\nonumber \\
Sp(2n+1):
&&  \,\,\,\,\, x_1,x_2,\ldots,x_n,\,{x}_1^{-1}, {x}_2^{-1},\ldots, {x}_n^{-1}, 1.
\nonumber
\end{eqnarray}
For example, in the case of $SO(2n)$ and $Sp(2n)$ we have $p_n(X, X^{-1}) = p_n(X)+p_n(X^{-1})$,
while for $SO(2n+1)$ and $Sp(2n+1)$,\, $p_n(X, X^{-1}, 1) = p_n(X)+p_n(X^{-1})+1$.
The connection to the group characters is obtained by inserting the eigenvalues into the Weyl
character formula and interpreting the exponetials as
\begin{equation}
e^{\lambda} = x_1^{\lambda_1}x_2^{\lambda_2}\ldots x_n^{\lambda_n},
\qquad\lambda=(\lambda_1,\ldots,\lambda_n).
\end{equation}
In the case of $U(n)$, the Weyl group is just the symmetric group (on $n$
letters). Hence the characters are labeled by partitions and the Weyl
character formula turns into the defining relation of the Schur functions.
Let $\mu$ be the conjugacy class of the permutation. One finds
\begin{equation}
\ch_\mu(\lambda) = \frac{\sum_{w\in S_n} \varepsilon(w)e^{w(\lambda+\rho).\mu}}
{\sum_{w\in S_n} \varepsilon(w)e^{w(\rho.\mu)}},
\end{equation}
where $\rho=(n-1,n-2,\ldots,1,0)$. Both numerator and denominator reduce to
determinants after inserting the $x_i$ (the denominator being the Vandermonde determinant),
and the quotient of the two alternating functions is a standard
construction of the Schur function.

Each  irreducible tensor representation, $V_{GL(n)}^\lambda$,
of $GL(n)$ is specified by a partition $\lambda$ of length $\ell(\lambda)\leq n$.
Let $X\in GL(n)$ have eigenvalues $(x_1,x_2,\ldots,x_n)$ and let $\rho=(n-1,n-2,\ldots,1,0)$.
Then the character of this irreducible representation is given by~\cite{Littlewood,Macdonald}:
$
{\rm ch}\, V_{GL(n)}^\lambda = a_{\lambda+\rho}(X)/a_\rho(X)
= \vert\,x_i^{\lambda_j+n-j}\,\vert/\vert\,x_i^{n-j}\,\vert=s_\lambda(X).
$
One can define the corresponding universal character
of $GL(n)$ by
\begin{equation}
{\rm ch}\, V_{GL}^\lambda = \{\lambda\}(X)=s_\lambda(X)\,,
\end{equation}
where $X =(x_1,x_2,\ldots)$. For each finite $n$ the characters ${\rm ch}\, V_{GL(n)}^\lambda$
are recovered from the universal characters ${\rm ch}\, V_{GL}^\lambda$ merely by setting
$X =(x_1,x_2,\ldots,x_n,0,0,\ldots,0)$.
In a similar way, there exists  irreducible tensor representation, $V_{O(n)}^\lambda$
and $V_{Sp(n)}^\lambda$, of $O(n)$ and $Sp(n)$, respectively. The corresponding
characters $\ch V_{O(n)}^\lambda$ and $\ch V_{Sp(n)}^\lambda$ may each be defined
in terms of determinants. More important is that there exist corresponding universal
characters \cite{Koike}, denoted by
\begin{equation}
{\rm ch}\, V_{O}^\lambda =
[\lambda](X)\quad\hbox{and}\quad {\rm ch}\, V_{Sp}^\lambda=\langle\lambda\rangle(X)\,,
\label{eq-universal-OSp}
\end{equation}
with $X =(x_1,x_2,\ldots)$ arbitrary. These are universal in the sense that for any finite $n$
the characters ${\rm ch}\, V_{O(n)}^\lambda$ and ${\rm ch} V_{Sp(n)}^\lambda$ are obtained by
specialising $X$ to $(x_1,x_2,\ldots,x_n,0,0,\ldots,0)$
with $x_1,x_2,\ldots,x_n$ restricted to the eigenvalues of the
appropriate group elements parametrised as above.
The universal characters (\ref{eq-universal-OSp}) are themselves defined by means
of the generating functions~\cite{Littlewood}:
\begin{eqnarray}
\prod_{i,j} (1-x_iy_j)^{-1}\prod_{i\leq j}(1-y_iy_j)
& = & \sum_\lambda [\lambda](X)\{\lambda\}(Y);
\\
\prod_{i,j} (1-x_iy_j)^{-1} \prod_{i < j} (1-y_iy_j)
& = & \sum_\lambda \langle \lambda\rangle(X)\{\lambda\}(Y)\,.
\end{eqnarray}
\begin{theorem} {\rm (Theorem 3.1 in \cite{King07})} The universal characters
${\rm ch}\, V_G^\lambda$ of the orthogonal and symplectic groups are given respectively by
\begin{eqnarray}
[\lambda](X) & = & \{\lambda/\textsf{C}\}(X) = s_{\lambda/\textsf{C}}(X), \,\,\,\,\,\,\,
{\rm where}
\,\,\,\,\, \textsf{C}(X) = \prod_{i\leq j}(1-x_ix_j);
\\
\langle\lambda\rangle(X) & = & \{\lambda/\textsf{A}\}(X) = s_{\lambda/\textsf{A}}(X),
\,\,\,\,\,\,\, {\rm where}
\,\,\,\,\, \textsf{A}(X) = \prod_{i<j}(1-x_ix_j).
\end{eqnarray}
\end{theorem}

{\bf Branching rules.}
In order to allow the possibility of extending the results to a wider class of subgroups of the general linear group, we consider any subgroup $H(n)$ of $GL(n)$, whose
group elements $X\in H(n)\subset GL(n)$ have the eigenvalues
$X =(x_1,x_2,\ldots,x_n)$. We assume, just as in the case of $O(n)$ and $Sp(n)$,
that there exist irreducible representations $V_{H(n)}^\lambda$
of $H(n)$, specified by partitions $\lambda$, with characters
$\ch V_{H(n)}^\lambda$ that may be determined by
specialising from $X =(x_1,x_2,\ldots)$ to $X =(x_1,x_2,\ldots,x_n,0,\ldots,0)$,
with appropriate $x_1,x_2,\ldots,x_n$.
Applying this to $O(n)$ and $Sp(n)$, we have:
\begin{theorem} {\rm (Theorem 4.1 in \cite{King07})} The branching rules for the decomposition of
representations of $GL(n)$ under restriction to the subgroups $O(n)$
and $Sp(n)$ take the form:
\begin{eqnarray}
GL(n)\supset O(n):\ && \{\lambda\}\rightarrow[\lambda/\textsf{D}]
\quad\hbox{with}\quad \textsf{D}= \textsf{C}^{-1}=\prod_{i\leq j} (1-x_ix_j)^{-1}\,;\\
GL(n)\supset Sp(n):\ && \{\lambda\}\rightarrow\langle\lambda/\textsf{B}\rangle
\quad\hbox{with}\quad \textsf{B}= \textsf{A}^{-1}=\prod_{i<j} (1-x_ix_j)^{-1}\,.
\end{eqnarray}
\end{theorem}
One of the most important features of group representations is the modular invariance of
their Ka\v{c}-Weyl character formula, which
allows us to derive many new results and to unify many
important results in topology.
Let us recall that the Weyl character formula in representation theory describes the
characters of irreducible representations of
compact Lie groups in terms of their highest weights.
\begin{remark}
All group like $\prod_j(1-F(x_j))^\alpha$, for polynomial $F$, {\rm (}based on 1-cocycles{\rm )}
induce trivial branchings, i.e. branchings equivalent to $U(n)$ {\rm (\cite{fauser:jarvis:2003a})}.
In the special case of the trivial one-dimensional representation the character is 1, so
the Weyl character formula becomes the Weyl denominator formula:
\begin{equation}
\sum_{w\in W} \varepsilon(w)e^{w(\rho)}
= e^{\rho}\prod_{\alpha \in \Delta_{+}}(1-e^{-\alpha}).\,
\end{equation}
For special unitary groups this is equivalent to the expression
$
\sum_{\sigma \in S_n} {\rm sgn}(\sigma) \, x_1^{\sigma(1)-1} \cdots x_n^{\sigma(n)-1}
\\
=\prod_{1\le i<j\le n} (x_j-x_i)
$
i.e. the Vandermonde determinant.
\end{remark}

{\bf Vertex operator traces.}
\label{Spectral2}
Vertex operators have played a fruitful role in string theory, mathematical constructions of
group representations as well as combinatorial constructions. We cite their applications to
affine and quantum affine Lie
algebras~\cite{Baker,lepowsky:wilson:1978a,frenkel:kac:1980a,frenkel:jing:1988a}
and sporadic discrete groups~\cite{frenkel:lepowsky:meurman:1988a}
(see also~\cite{kac:1990a}, Chapter 14).
Variations on the theme of symmetric
functions are applications, for example, to $Q$-functions~\cite{jing:1991a,salam:wybourne:1992a},
Hall-Littlewood functions~\cite{jing:1991b}, Macdonald
functions~\cite{jing:jozefiak:1992a,etingof:kirillov:1995a},
Jack functions~\cite{cai:jing:2010a}, Kerov's symmetric functions  \cite{Kerov}
(and a specialization of $S$-functions introduced by Kerov).

By considering different specializations of Kerov's symmetric
functions the trace calculations in representations of the levels
quantum affine algebra $U_q({\mathfrak g}{\mathfrak l}_N)$ can be
feasible. The extension of this mathematical tools to other
(quantum) affine algebras and superalgebras is also practicable
and provides the relevant vertex operator realizations of those
algebras.

As an approach to generalising the vertex operators, the observations made in
previous sections allow us to write down the expressions for replicated or parameterized vertex
operators. In the simplest case, this is exemplified by
\begin{equation}
V_\alpha(z) = M(\alpha z)\, L^\perp(\alpha z^{-1})
= \exp\left(  \alpha \sum_{k\geq1}\frac{z^k}{k}\, p_{k}\,\right)
\exp  \left(-\alpha\sum_{k\geq1}\, z^{-k}\, \frac{\partial}{\partial p_k}
\right),
\label{Eq-VOalpha}
\end{equation}
for any $\alpha$, integer, rational, real or complex. Then we have:
\begin{eqnarray}
 M(\alpha z;X) & = &  M(z;X)^\alpha = \prod_{i\geq1} (1-z\,x_i)^{-\alpha}
= \sum_{\sigma}\, s_\sigma(\alpha z)\, s_\sigma(X)
\nonumber \\
& = &  \sum_{\sigma}\, z^{|\sigma|} \dim_\sigma(\alpha)\ s_\sigma(X)\,,
\\
L(\alpha z^{-1};X)
& = & L(z^{-1};X)^\alpha = \prod_{i\geq1} (1-z^{-1}\,x_i)^\alpha
 = \sum_{\tau}\, (-1)^{|\tau|}\, s_{\tau}(\alpha z^{-1})\, s_{\tau'}(X)
   \nonumber \\
  & = & \sum_{\tau}\, (-z)^{-|\tau|} \dim_{\tau}(\alpha)\ s_{\tau'}(X)\,,
\end{eqnarray}
as given first in~\cite{jarvis:yung:1993a}. The following result holds
\begin{equation}
V_{\alpha_1}(x_1)\cdots V_{\alpha_n}(x_n) = \prod_{i< j}(1-x_jx_i^{-1})^{\alpha_i\alpha_j}
: V_{\alpha_1}(x_1)\cdots V_{\alpha_n}(x_n) :
\end{equation}
Here by $:\, :$ we mean the procedure normal ordering.
The Cauchy kernel $M(XZ)$ serves as a generating function for characters of $GL(n)$ in the sense
that
\begin{equation}
M(XY) = \prod_{i,j}(1-x_iy_j) = \sum_\lambda s_\lambda(X)s_\lambda(Y),
\end{equation}
where $s_\lambda(X)$ is the character of the irreducible representation $V_{GL(n)}^\lambda$
of highest weight $\lambda$ evaluated at group elements whose eigenvalues are the element of $X$.
We summarize some useful formulas:
\begin{eqnarray}
M(q; XY) & = &  \prod_{i,j}(1-qx_iy_j)^{-1}  =  \sum_\alpha q^\alpha s_\alpha(X)s_\alpha(Y),
\nonumber \\
L(q; XY) & = & \prod_{i,j}(1-qx_iy_j) \,\,\,\, =
\,\,\, \sum_\alpha (-q)^{|\alpha|} s_\alpha(X)s_{\alpha^\prime}(Y).
\end{eqnarray}

The standard infinite symmetric function series which used in
representation theory of finite dimensional algebras and
superalgebras can be given by a Fock space interpretation \cite{King90}
\begin{eqnarray}
{}_q\langle \textsf{A}\vert : V_{1}(x_1)\cdot V_{2}(x_2) \cdots  : \vert 0\rangle & = &
\prod_{i<j}(1-qx_ix_j) = \sum_{\alpha \in \textsf{A}}(-q)^{|\alpha|/2}s_\alpha(X)\,,
\\
{}_q\langle \textsf{B}\vert : V_{1}(x_1)\cdot V_{2}(x_2) \cdots  : \vert 0\rangle & = &
\prod_{i<j}(1-qx_ix_j)^{-1} = \sum_{\beta \in \textsf{B}} q^{|\beta|/2}s_\beta(X)\,,
\\
{}_q\langle \textsf{C}\vert : V_{1}(x_1)\cdot V_{2}(x_2) \cdots  : \vert 0\rangle & = &
\prod_{i\leq j}(1-qx_ix_j) = \sum_{\gamma \in \textsf{C}}(-q)^{|\gamma|/2}s_\gamma(X)\,,
\\
{}_q\langle \textsf{D}\vert : V_{1}(x_1)\cdot V_{2}(x_2) \cdots  : \vert 0\rangle & = &
\prod_{i\leq j}(1-qx_ix_j)^{-1} = \sum_{\delta \in \textsf{D}} q^{|\delta|/2}s_\delta(X)\,.
\end{eqnarray}
To define $\textsf{A},\, \textsf{B},\, \textsf{C},\, \textsf{D}$ is is convenient to use
the Frobenius notation:
\begin{eqnarray}
\mathfrak{F}_n\ = \left\{
\left(\begin{array}{cccc}
    a_1 & a_2 & \ldots & a_r \\
    b_1 & b_2 & \ldots & b_r
    \end{array}\right) \in \mathfrak{F}\ \,\bigg|\,   a_k-b_k=n
    \quad \hbox{for all}\quad
    \begin{array}{l} r=0,1,2,\ldots \\ k=1,2,\ldots,r\\ \end{array} \right\}
\label{Eq-frobenius-sets}
\end{eqnarray}
With this notation $\textsf{A} = \mathfrak{F}_{-1},\,\,
\textsf{C}= \mathfrak{F}_1$,\, $\textsf{D}$ is the set of
partitions all of whose parts are even, and $\textsf{B}$ is the
set of partitions all of whose distinct parts are repeated an even
number of times. Let us define modified symmetric functions (and
corresponding kets vectors): $S_{\lambda/\textsf{C}}(X),\,\,
\langle\lambda/\textsf{C}\vert,\,\, S_{\lambda/\textsf{A}}(X),\,\,
\langle\lambda/\textsf{A}\vert$, where the notation indicates
symmetric function division distributed over all admissible
elements of the indicated series. Then matrix elements of vertex
operators with the corresponding reservoir states
\begin{eqnarray}
{}_q\langle \textsf{A}/\textsf{C}\vert : V_{1}(x_1)\cdot V_{2}(x_2) \cdots  : \vert 0\rangle
& = &
\prod_{k=1}^\infty\prod_{i<j}(1-q^kx_ix_j)
\nonumber \\
&= & \prod_{i<j} {\mathcal R}(s = (1+\Omega(x_ix_j; \vartheta))(1-i\varrho(\vartheta)))\,,
\\
{}_q\langle \textsf{C}/\textsf{A}\vert : V_{1}(x_1)\cdot V_{2}(x_2) \cdots  : \vert 0\rangle
& = &
\prod_{k=1}^\infty\prod_{i\leq j}(1-q^kx_ix_j)
\nonumber \\
& = &
\prod_{i\leq j}{\mathcal R}(s = (1+\Omega(x_ix_j; \vartheta))(1-i\varrho(\vartheta)))\,,
\end{eqnarray}
where $\Omega(x_ix_j; \vartheta)  :=  {\rm log}(x_ix_j)/ 2\pi i \vartheta$.

\section{The quantum group invariants}
\label{Quantum}

\subsection{The HOMFLY skein and the quantum group invariants}
\label{HOMFLY}

{\bf Preliminaries.}
Our notations are summarized as follows:
denote by $\cY$ the set of all Young diagrams. Let $\chi_A$ be the character of irreducible
representation of symmetric group, labelled by partition $A$. Given a partition $\mu$,
define $m_j = \textrm{card} (\mu_k=j; k\geq 1)$. The order of the conjugate class of type
$\mu$ is given as before by: $\mathfrak{z}_\mu = \prod_{j\geq1} j^{m_j} m_j!.$
The symmetric power functions of a given set of variables $X=\{x_j\}_{j\geq 1}$
are defined as the direct limit of the Newton polynomials:
$p_n(X) = \sum_{j\geq1} x_j^n, \, \, p_\mu(X) = \prod_{i\geq 1} p_{\mu_i}(X),$
and we have the following formulae which determines the Schur function and the orthogonality
property of the character
\begin{equation}
s_A(X) = \sum_{\mu} \frac{ \chi_A(C_\mu) }{\mathfrak{z}_\mu} p_\mu(X), \,\,\,\,\,\,\,\,\,
\sum_{\mu} \frac{ \chi_A(C_\mu) \chi_B(C_\mu) }{ \mathfrak{z}_\mu } = \delta_{A,B}\,.
\end{equation}
Given $X= \{x_i\}_{i\geq 1}$, $Y=\{y_j\}_{j\geq 1}$, define
$
X\ast Y = \{x_i\cdot y_j\}_{i\geq 1, j\geq 1}.
$
We also define $X^d = \{ x_i^d\}_{i\geq 1}$.
The $d$-th Adam operation of a Schur function is given by $s_A(X^d)$.
We use the following convention for the notations:

-- Denote by $\cL$ a link and by $L$ the number of components in $\cL$.

-- The irreducible $U_q(\mathfrak{sl}_N)$ module associated to $\mathcal L$ will be labelled by
their highest   weights, thus by Young diagrams. We usually denote it by a vector form
$\overrightarrow{A}=(A^1,\ldots,A^L)$.

-- Let $\overrightarrow{X} =(x_1,\ldots,x_L)$ is $L$ sets of variables,
each of which is associated to  a component of $\mathcal L$ and
$\overrightarrow{\mu} = (\mu^1,\ldots,\mu^L)\in\cY^L$ be a tuple of $L$ partitions, and
$$
[\overrightarrow{\mu}] = \prod_{\alpha=1}^L [\mu^\alpha],
\,\,\,\,\,\,\, \mathfrak{z}_{\overrightarrow{\mu}} = \prod_{\alpha=1}^L \mathfrak{z}_{\mu^\alpha},
\,\,\,\,\,\,\, \chi_{\overrightarrow{A}}(C_{\overrightarrow{\mu}}) =
\prod_{\alpha=1}^L \chi_{A^\alpha}(C_{\mu^\alpha}),
$$
$$
s_{\overrightarrow{A}}(\overrightarrow{X}) =
\prod_{\alpha=1}^L s_{A^\alpha}(x_\alpha), \,\,\,\,\,\,\,
p_{\mu }(X)=\overset{\ell (\mu )}{\prod_{i=1}}p_{\mu _{i}}(X),\,\,\,\,\,\,\,
p_{\overrightarrow{\mu}}(\overrightarrow{X}) = \prod_{\alpha=1}^L p_{\mu^\alpha}(x_\alpha).
\nonumber
$$

The quantum group invariants can be defined over any semi-simple Lie algebra $\mathfrak g$.
In the $SU(N)$ Chern-Simons gauge
theory we study the quantum ${\mathfrak s}{\mathfrak l}_N$ invariants, which can be identified as
the colored HOMFLY polynomials.

The framed HOMFLY polynomial of links (an invariant of framed oriented links), denote
by $\cH(\cL)$, and can be normalize as:
$\cH(\bigcirc) = (t^{-\frac12}-t^{\frac12})/(q^{-\frac12}-q^{\frac12})$.
These invariants can be recursively computed through the HOMFLY skein.

The colored HOMFLY polynomials are defined through \emph{satellite knot}. A satellite of $\cK$ is
determined by choosing a diagram $Q$ in the annulus. Draw $Q$ on the annular neighborhood of
$\cK$ determined by the framing to give a satellite knot $\cK\star Q$.
One can refer to this construction as \emph{decorating $\cK$ with the pattern $Q$}. The HOMFLY
polynomial $\cH(\cK\star Q)$ of the satellite depends on $Q$ only as an element of the skein
$\cC$ of the annulus. $\{Q_\lambda\}_{\lambda\in\cY}$ form a basis of $\cC$. $\cC$ can be
regarded as the parameter space for these invariants of $\cK$, and can be  called as
the HOMFLY {\it satellite invariants of $\cK$}.

\subsection{Link invariants from vertex models}

Vertex calculation (see Sect. \ref{Hopf}) can be used to associate oriented graphs, provided one
decides on a way
of numbering the edges which meet at a given vertex. It is known that plan projection of any
link is a valent graph and therefore the partition function associates a number to every
oriented link.
One can choose the weights of verticies (beyond the quantum Yang-Baxter equation) in such
a way that the partition
function depends only on the equivalent class of the link.

The quantum $\mathfrak{sl}_N$ invariant for the irreducible module $V_{A^1},\ldots,V_{A^L}$,
labeled by the corresponding partitions $A^1,\ldots, A^L$,  can be
identified as the HOMFLY invariants for the link decorated by $Q_{A^1},\ldots,Q_{A^L}$.
The quantum $\mathfrak{sl}_N$ invariants of the link is given by
\begin{equation}
W_{\overrightarrow{A}}(\mathcal{L}; q,t) =
\mathcal{H} (\mathcal{L}\star \otimes_{\alpha=1}^L Q_{A^\alpha} )\,.
\label{W}
\end{equation}
The colored HOMFLY polynomial of the link $\mathcal L$ (\ref{W}) can be defined by \cite{Zhu}
\begin{equation}
W_{\overrightarrow{A}} = q^{-\sum_{\alpha = 1}^Lk_{A^\alpha}\omega({\mathcal K}_\alpha)}
t^{-\sum_{\alpha = 1}^L \vert A^\alpha\vert \omega({\mathcal K}_\alpha)}
\langle \mathcal{L}\star \otimes_{\alpha=1}^L Q_{A^\alpha} \rangle\,,
\end{equation}
where $\omega({\mathcal K}_\alpha)$ is the number of the $\alpha$-component
${\mathcal K}_\alpha$ of $\mathcal L$
and the bracket $\langle \mathcal{L}\star \otimes_{\alpha=1}^L Q_{A^\alpha} \rangle$ denotes
the framed HOMFLY polynomial of the satellite link
$\mathcal{L}\star \otimes_{\alpha=1}^L Q_{A^\alpha}$.
We can define the following invariants:
\begin{equation}\label{definition of Z_mu}
\textsf{Z}_{\overrightarrow{\mu}}(\mathcal{L};q,t) =
\sum_{\overrightarrow{A} =(A^1,\ldots,A^L) } \bigg( \prod_{\alpha=1}^L
\chi_{A^\alpha}(C_{\mu^\alpha} ) \bigg)
W_{\overrightarrow{A}}(\mathcal{L};q,t)\,.
\end{equation}

The Chern-Simons partition function $\textsf{Z}_{CS}({\mathcal L};q,t)$ and the free energy
$F({\mathcal L};q,t)$ of the link ${\mathcal L}$ are the following generating series of
quantum group invariants weighted by Schur functions and by the
invariants $Z_{\overrightarrow{\mu}}$:
\begin{eqnarray}
\textsf{Z}_{CS}({\mathcal L};q,t) & = &
1+ \sum_{\overrightarrow{A}} W_{\overrightarrow{A}} ({\mathcal L}; q,t)
s_{\overrightarrow{A}}(\overrightarrow{X}) =
1+ \sum_{\overrightarrow{\mu} }
\frac{ \textsf{Z}_{\overrightarrow{\mu}}
({\mathcal L};q,t) }{{\mathfrak z}_{\overrightarrow{\mu}} }
p_{\overrightarrow{\mu}}(\overrightarrow{X}) \,,
\label{CS-A}
\\
F({\mathcal L};q,t) & = & \log Z_{CS}({\mathcal L};q,t)
= \sum_{\overrightarrow{\mu}}
\frac{ F_{\overrightarrow{\mu}}({\mathcal L};q,t) }{{\mathfrak z}_{\overrightarrow{\mu}}}
p_{\overrightarrow{\mu}}(\overrightarrow{X}) \,.
\end{eqnarray}

\subsection{From summations to infinite products}
\label{Product}

The Chern-Simons theory has been conjectured to be equivalent to a topological string theory
$1/N$ expansion in physics. This duality conjecture builds a fundamental connection in
mathematics. On the one hand, Chern-Simons theory leads to the
construction of knot invariants; on the other hand, topological string theory gives rise to
Gromov-Witten theory in geometry.

The Chern-Simons/topological string duality conjecture identifies
the generating function of Gromov-Witten invariants as
Chern-Simons knot invariants \cite{OV}.  Based on these thoughts,
the existence of a sequence of integer invariants is conjectured
\cite{OV, LMV} in a similar spirit to  Gopakumar-Vafa
setting \cite{GV}, which provides an essential evidence of the
duality between Chern-Simons theory and topological string theory.
This integrality conjecture is called the LMOV
conjecture. One important corollary of the LMOV conjecture is to
express Chern-Simons partition function as an infinite product
derived in this article. The motivation of studying such an
infinite-product formula is based on a guess on the modularity
property of topological string partition function.

To derive an infinite-product formula, we will state the result
for a knot at first, since the notations in the computation for a
knot are relatively simpler.

{\bf The case of a knot.}
Based on LMOV conjecture the following infinite product for a knot has been obntained
in \cite{LiuPeng}:
\begin{equation}
\textsf{Z}_{CS}^{SL}(\cK;q,t;X)
=   \prod_{\mu}
\prod_{Q\in {\mathbb Z}/2} \,\,
\prod_{m=1}^\infty \;
\prod_{k = -\infty}^\infty\;
\big \langle 1- q^{k+m}t^Q   X^\mu
\big \rangle^{-m\, {n}_{\mu;\,g,Q}}\,.
\end{equation}
Here ${n}_{\mu;\,g,Q}$ are invariants related to the integer invariants in the LMOV
conjecture. For a given $\mu$, ${n}_{\mu;g,Q}$ vanish for sufficiently large
$|Q|$ due to the vanishing property of $n_{B;\,g,Q}$; the products involved with $Q$ and $k$
are finite products for a fixed partition $\mu$. The symmetric product
$\langle\,\cdot\,\rangle$ defined by the formula
\begin{equation}
\label{SP}
\big\langle 1- \psi X^\mu \big\rangle
= \prod_{ x_{i_1},\ldots, x_{i_{\ell(\mu)} } }
\Big( 1- \psi x_{i_1}^{\mu_1} \cdots x_{i_{\ell(\mu)}}^{\mu_{\ell(\mu)}} \Big).
\end{equation}
\begin{Remark}
The symmetric product {\rm(\ref{SP})} can be simplify by using of the Bell polynomials
{\rm(\ref{P1})}, {\rm(\ref{Q1})}, Sect. \ref{Mult}. In that case the final result can be
represent in form similar to {\rm(\ref{F1})} and {\rm(\ref{G1})}.
\end{Remark}

In terms of Ruelle spectral functions $\textsf{Z}_{CS}^{SL}(\cK;q,t;X)$ takes the form
\begin{eqnarray}
&&
\textsf{Z}_{CS}^{SL}(\cK;q,t;X) =
\prod_{\mu}\prod_{Q\in {\mathbb Z}/2} \,\prod_{k = -\infty}^\infty \,\prod_{m=1}^\infty
\prod_{ x_{i_1},\ldots, x_{i_{\ell(\mu)} } }
\Big( 1- q^{k+m}t^Q x_{i_1}^{\mu_1}
\cdots x_{i_{\ell(\mu)}}^{\mu_{\ell(\mu)}} \Big)^{{-m\, {n}_{\mu;\,g,Q}}}
\nonumber \\
= &&
\prod_{\mu}\prod_{Q\in {\mathbb Z}/2} \,\prod_{k = -\infty}^\infty \,\prod_{m=1}^\infty
\prod_{ x_{i_1},\ldots, x_{i_{\ell(\mu)} } }
\Big( 1- q^{m + \Omega(q^kt^QX^\mu; \vartheta)} \Big)^{-m\, {n}_{\mu;\,g,Q}}
\nonumber \\
= &&
\prod_{\mu}\prod_{Q\in {\mathbb Z}/2} \,
\prod_{k = -\infty}^\infty \,\prod_{ x_{i_1},\ldots, x_{i_{\ell(\mu)} }}\, \prod_{m=1}^\infty
\left({\mathcal R}(s= (m+\Omega(t^QX^\mu q^k;\vartheta))
(1-i\varrho(\vartheta)))\right)^{-{n}_{\mu;\,g,Q}},
\label{Knot1}
\end{eqnarray}
where as before $\Omega(q^kt^QX^\mu; \vartheta)
\equiv {\rm log}(q^kt^Q x_{i_1}^{\mu_1}
\cdots x_{i_{\ell(\mu)}}^{\mu_{\ell(\mu)}})/2\pi i\vartheta$.

{\bf The case of links.}
The generalization of this result for the case of links can be easy derived.
Let $\overrightarrow{\mu}=(\mu^1,\ldots,\mu^L)$ and $\overrightarrow{X}=(x_1,\ldots, x_L)$.
Denote by $\ell_i$ the length of $\mu^i$. Generalize the symmetric product
in Eq. (\ref{SP}) to $\overrightarrow{\mu}$ and $\overrightarrow{X}$ as:
\begin{equation}
\big\langle 1-\psi\, \overrightarrow{X}^{{\mu}}\big\rangle
=   \prod_{\alpha=1}^L\
\prod_{i_{\alpha, 1},\ldots, i_{\alpha, \ell_\alpha} }
\Big( 1- \psi\prod_{\alpha=1}^L
\big(  (x_\alpha)_{ i_{\alpha,1} }^{\mu^\alpha_1} \cdots
(x_\alpha)_{ i_{\alpha, \ell_\alpha} }^{\mu^\alpha_{\ell_\alpha}}\big)\Big)\,.
\end{equation}
The infinite-product formula for the Chern-Simons partition function of ${\mathcal L}$:
\begin{eqnarray}
\!\!\!\!\!\!\!\!\!\!\!\!\!\!\!\!\!\!\!\!\!
&&
\textsf{Z}_{CS}^{SL}({\mathcal L};q,t; \overrightarrow{X}) =
\nonumber \\
\!\!\!\!\!\!\!\!\!\!\!\!\!\!\!\!\!\!\!\!\!
&&
\prod_{\overrightarrow{\mu}}\,
\prod_{Q\in {\mathbb Z}/2} \,\prod_{k= -\infty}^\infty\,\prod_{m=1}^\infty
\prod_{\alpha=1}^L\
\prod_{i_{\alpha, 1},\ldots, i_{\alpha, \ell_\alpha} }
\!\!\!\!\Big(1- q^{m+g-2k}
\prod_{\alpha=1}^L
\big(  (x_\alpha)_{ i_{\alpha,1} }^{\mu^\alpha_1} \cdots
(x_\alpha)_{ i_{\alpha, \ell_\alpha} }^{\mu^\alpha_{\ell_\alpha}}\big)\Big)^{-m\, {n}_{\overrightarrow{\mu};\,g,Q}} =
\nonumber \\
\!\!\!\!\!\!\!\!\!\!\!\!\!\!\!\!\!\!\!\!\!
&&
\prod_{\overrightarrow{\mu}}\,
\prod_{Q\in {\mathbb Z}/2} \,\prod_{k= -\infty}^\infty\,\prod_{\alpha=1}^L\
\prod_{i_{\alpha, 1},\ldots, i_{\alpha, \ell_\alpha} }\prod_{m=1}^\infty
{\mathcal R}\left(s= (m+\Omega(q^{g-2k}t^Q\overrightarrow{X}^\mu; \vartheta))
(1-i\varrho(\vartheta))\right)^{-n_{\overrightarrow{\mu};\,g,Q}}\!\!\!\!\!\!\!\!\!\!.
\end{eqnarray}
\\

{\bf The case of the unknot.}
The Chern-Simons partition function of the unknot is given by
\begin{equation}
\textsf{Z}_{CS}^{SL}(\bigcirc;\, q,t)
= 1+ \sum_{A} {\rm dim}_q V_A \cdot s_A(X),
\label{unknot}
\end{equation}
where ${\rm dim}_q V_A$ is the quantum dimension of the irreducible
$U_q({{\mathfrak s}{\mathfrak l}}_N)$ module $V_A$. The formula of
quantum dimension is well known (see for example
\cite{Liu2}):
\begin{equation}
{\rm dim}_q V_A
= \sum_\mu \frac{ \chi_A(C_\mu) }{ {\mathfrak z}_\mu } \prod_{j=1}^{\ell(\mu)}
\frac{ t^{-\frac{ \mu_j}{2}} - t^{\frac{ \mu_j}{2}} }
{ q^{-\frac{ \mu_j}{2} } - q^{\frac{ \mu_j}{2}}}\,.
\end{equation}
Then a similar computation leads to the following formula:
\begin{eqnarray}
\label{infinite-product formula for unknot}
\textsf{Z}_{CS}^{SL}(\bigcirc;\, q,t; X) & = & \prod_{m=1}^\infty \prod_{i}\frac{(1- q^m t^{1/2}x_i)^m}{(1- q^m t^{-1/2}x_i)^m}
= \prod_{m=1}^\infty \prod_{i}\frac{( 1- q^{m +\Omega(t^{1/2}x_i; \vartheta)})^m}
{(1- q^{m +\Omega(t^{-1/2}x_i; \vartheta)}))^m}
\nonumber \\
& = &
\prod_{i}\prod_{m=1}^\infty
\frac{{\mathcal R}(s=(m+\Omega(t^{1/2}x_i; \vartheta))(1-i\varrho(\vartheta)))}
{{\mathcal R}(s=(m+\Omega(t^{-1/2}x_i; \vartheta))(1-i\varrho(\vartheta)))}\,.
\end{eqnarray}

\subsection{Singularities and symmetries in infinite-product structure}
\label{Symmetry}

In this section we discuss a basic symmetric property of infinite-product structure
obtained from the LMOV partition function. First we recall
some results on the Ruelle (Patterson-Selberg type) spectral functions. For details
we refer the reader to \cite{BB,BCST} where spectral
functions of hyperbolic three-geometry were considered in connection with
three-dimensional Euclidean black holes, pure supergravity, and string amplitudes.

Let ${\Gamma}^\gamma \in G=SL(2, {\mathbb C})$ be the discrete group defined by
\begin{eqnarray}
{\Gamma}^\gamma & = & \{{\rm diag}(e^{2n\pi ({\rm Im}\,\vartheta + i{\rm
Re}\,\vartheta)},\,\,  e^{-2n\pi ({\rm Im}\,\vartheta + i{\rm Re}\,\vartheta)}):
n\in {\mathbb Z}\} = \{{\gamma}^n:\, n\in {\mathbb Z}\}\,,
\nonumber \\
{\gamma} & = & {\rm diag}(e^{2\pi ({\rm Im}\,\vartheta + i{\rm
Re}\,\vartheta)},\,\,  e^{-2\pi ({\rm Im}\,\vartheta + i{\rm Re}\,\vartheta)})\,.
\label{group}
\end{eqnarray}
One can construct a zeta function of Selberg-type for the group
${\Gamma}^\gamma \equiv {\Gamma}_{(\alpha, \beta)}^\gamma$ generated by a
single hyperbolic
element of the form ${\gamma_{(\alpha, \beta)}} = {\rm diag}(e^z, e^{-z})$,
where $z= \alpha +i\beta$ for $\alpha, \beta >0$. Actually $\alpha = 2\pi {\rm Im}\,\vartheta$
and $\beta = 2\pi {\rm Re}\,\vartheta$.
The Patterson-Selberg spectral function $Z_{{\Gamma}^\gamma} (s)$ and its logarithm
for ${\rm Re}\, s> 0$ can be attached
to $H^3/{\Gamma}^\gamma$ as follows:
\begin{eqnarray}
Z_{{\Gamma}^\gamma}(s) & := & \prod_{k_1,k_2\geq
0}[1-(e^{i\beta})^{k_1}(e^{-i\beta})^{k_2}e^{-(k_1+k_2+s)\alpha}]\,,
\label{zeta00}
\\
{\rm log}\, Z_{{\Gamma}^\gamma} (s) \!\! & = &
-\frac{1}{4}\sum_{n = 1}^{\infty}\frac{e^{-n\alpha(s-1)}}
{n[\sinh^2\left(\frac{\alpha n}{2}\right)
+\sin^2\left(\frac{\beta n}{2}\right)]}\,.
\label{logZ}
\end{eqnarray}
The zeros of $Z_{\mG^\gamma} (s)$ are precisely the set of complex numbers
\begin{equation}
\zeta_{n,k_{1},k_{2}} = -\left(k_{1}+k_{2}\right)+ i\left(k_{1}-
k_{2}\right) \beta/\alpha + 2\pi i n/\alpha\,,
\label{zeroes}
\end{equation}
with $n \in {\mathbb Z}$.
The Ruelle functions $\cR(s)$, (\ref{R1}) -- (\ref{RU2}), are an alternating product of factors,
each of which is a Selberg-type zeta function; $\cR(s)$ can be continued meromorphically
to the entire complex plane $\mathbb C$. For more information about the analytic properties
of this spectral function we refer the reader to the papers \cite{Patterson,BCST}.
The magnitude of the zeta-function is bounded
for both ${\rm Re}\,s\geq 0$ and ${\rm Re}\,s\leq 0$, and its growth can be estimated as
\begin{equation}
\big|Z_{\mG^\gamma}(s) \big| \leq \Big(\,
\prod_{k_1+k_2\leq
|s|}\,  \e^{|s|\, \ell}\, \Big)\,
\Big(\, \prod_{k_1+k_2\geq
|s|}\, \big(1- \e^{(|s|-k_1-k_2)\, \ell} \big)\,\Big)
\leq C_1\,\e^{C_2\, |s|^3}
\label{estimate}
\end{equation}
for suitable constants $\ell,C_1, C_2$. The first product on the right-hand side of
(\ref{estimate}) gives the exponential growth, while the second product is bounded.
The spectral function $Z_{\mG^\gamma} (s)$ is an entire function of
order three and of finite type which can be written as a Hadamard product~\cite{BCST}
\begin{equation}
Z_{\mG^\gamma}(s) =
\e^{Q(s)} \
\prod_{\zeta \in {\Sigma}}\,
\Big(\, 1-\frac{s}{\zeta}\, \Big)\,
\exp \Big(\,
\frac{s}{\zeta} + \frac{s^2}{2\zeta^2} +
\frac{s^3}{3\zeta^3}\, \Big)\ ,
\label{Hadamard}
\end{equation}
where $\Sigma$ is the set of zeroes $\zeta := \zeta_{n,k_{1},k_{2}}$
and $Q(s)$ is a polynomial of degree at most three. (The product formula for entire
function (\ref{Hadamard}) is also known as Weierstrass formula  (1876).)

As a function of $\Omega(\cdots ; \vartheta)$ the partition function $\textsf{Z}_{CS}$
has infinitely many poles of orders $m\in {\mathbb Z}_+$ (in fact the functions ${\mathcal R}(s)$
are the meromorphic functions; poles of ${\mathcal R}(s)$ correspond to zeros of
$Z_{\mG^\gamma}(s)$).

{\bf Symmetry property.} Let us discuss symmetry properties of the infinite-product formula
given in Sect. \ref{Product}. For this reason we can use functional equations for the
spectral Ruelle functions (\ref{R1})-- (\ref{RU2}):
\begin{eqnarray}
\!\!\!\!\!\!\!\!\!\!
&&
\!\!\!\!\!\!\!\!\!\!\!\!\!\!
{\mathcal R}(s= ({z}+b)(1-i\varrho(\vartheta))+i\sigma(\vartheta))\cdot
{\mathcal R}(s= -(1+{z}+b)(1-i\varrho(\vartheta))+i\sigma(\vartheta))
\nonumber \\
\!\!\!\!\!\!\!\!\!\!
=
&&
\!\!\!\!\!\!\!
q^{-{z}b-b(b+1)/2}
{\mathcal R}(s= -{z}(1-i\varrho(\vartheta))+i\sigma(\vartheta))\cdot
{\mathcal R}(s= (1+{z})(1-i\varrho(\vartheta))+i\sigma(\vartheta))
\nonumber \\
\!\!\!\!\!\!\!\!\!\!
=
&&
\!\!\!\!\!\!\!
q^{-{z}(b-1)-b(b+1)/2}
{\mathcal R}(s= (1-{z})(1-i\varrho(\vartheta))+i\sigma(\vartheta))\cdot
{\mathcal R}(s= {z}(1-i\varrho(\vartheta))+i\sigma(\vartheta)).
\label{DE3}
\end{eqnarray}
The first key is Eq. (\ref{Knot1}) (the case of links is exactly similar).
The simple case $b=0$ in Eq. (\ref{DE3}) leads to the symmetry
$\vartheta \rightarrow -\vartheta$, i.e the symmetry $q\rightarrow q^{-1}$.

There is also the following symmetry about $\mu$ and $Q$
$
n_{\mu;\, g,-Q} = (-1)^{\ell(\mu)} n_{\mu;\, g, Q},
$
which can be interpreted as the rank-level duality of the $SU(N)_k$ and $SU(k)_N$
Chern-Simons gauge theories \cite{LiuPeng}. Rank-level duality is essentially a symmetry of
quantum group invariants relating a labeling color to its transpose \cite{LiuPeng}. It can be
expressed using symmetry about $\mu$, $Q$, and modularity properties of Ruelle functions
as follows:
$
W_{ A^t}( s^{-1}, -v ) = W_{A} (s, v)\,,
$
where $s=q^{1/2}$, $v=t^{1/2}$. The stronger version is \cite{LiuPeng,Zhu,CLPZ}:
$
W_{A^t}( s^{-1}, v) = (-1)^{|A|} W_A (s, v),\,
W_A(s, -v) = (-1)^{|A|} W_A(s, v).
$

\section{Orthogonal group and colored Kauffman polynomials}
\label{Orthogonal}


{\bf Quantum Invariants of Links.}
Recall that for the unknot $\bigcirc $, $W_{A}(\bigcirc )$ is the
quantum dimension $\dim _{q}(V_{A})$ of the corresponding representation
space $V_{A}$ (see Eq. (\ref{unknot})).

-- If $\mathfrak{g}= {\mathfrak s}{\mathfrak l}_{N}$ and $A^{1}=A^{2}=\cdot \cdot \cdot
=A^{L}=(1)$, the quantum group invariant of links equal to the HOMFLY
polynomial at $t=q^{N}$ up to a universal factor $(t-t^{-1})/(q-q^{-1})$.

-- If $\mathfrak{g}=so_{2N+1}$ and $A^{1}=A^{2}=\cdot \cdot
\cdot =A^{L}=(1)$, quantum group invariant of links equal to Kauffman
polynomial at $t=q^{2N}$ up to a universal factor $1+ (t-t^{-1})/(q-q^{-1})$
and some $t$ power of the linking numbers.

The quantum group invariant associated to $\mathfrak{g}= {\mathfrak s}{\mathfrak l}_{N}$ and
$\mathfrak{g}=so_{2N+1}$ are called the colored HOMFLY and the colored
Kauffman polynomials respectively.
Actually the irreducible representation of the quantum groups of special
linear and orthogonal cases can be labeled by the Young Tableau.

For each link $\mathcal{L}$, the type$-A$ Chern-Simons partition function of
$\mathcal{L}$ is defined by (\ref{CS-A}).
The original LMOV conjecture describes a very subtle structure of
$\textsf{Z}_{CS}^{SL}(\mathcal{L};q,t;\overrightarrow{X})$, which was proved in
\cite{Liu2}, based on the cabling technique and a
careful degree analysis of the cut-join equation. As an application, the
LMOV conjecture gives highly non-trivial relations between colored HOMFLY
polynomials. The first such relation is the classical Lichorish-Millett
theorem \cite{LiM}.

The study of the colored Kauffman polynomials are more difficult. For
instance, the definition of the Chern-Simons partition function for the
orthogonal quantum groups involves the representations of Brauer centralizer
algebras, which admit a more complicated orthogonal relations \cite{Ram1,
Ram2,Ram3}.
The orthogonal quantum group version of LMOV conjecture has been formulated in
\cite{C-C} by using the representation of the Brauer centralizer algebra.

Let $\textsf{Z}_{CS}^{SO}(\mathcal{L},q,t)$ be the orthogonal Chern-Simons partition
function defined by
\begin{equation}
\textsf{Z}_{CS}^{SO}(\mathcal{L};q,t;\overrightarrow{X})=\sum_{\overrightarrow{\mu }
\in \mathcal{P}^{L}}\frac{p_{\overrightarrow{\mu }}(\overrightarrow{X})}{
\mathfrak{z}_{\overrightarrow{\mu}}}\underset{\overrightarrow{A}\in \widehat{Br}_{|
\overrightarrow{\mu }|}}{\sum }\chi _{\overrightarrow{A}}(\gamma _{
\overrightarrow{\mu }})W_{\overrightarrow{A}}^{SO}(\mathcal{L};q,t),
\end{equation}
where $\widehat{Br}_{|\overrightarrow{\mu }|}$ denotes the set $\widehat{Br}
_{d^{1}}\times \cdot \cdot \cdot \times \widehat{Br}_{d^{L}}$ (every element
is a representation of the Brauer algebra), $\overrightarrow{\mu }=(\mu
^{1},...,\mu ^{L})$ for partitions $\mu ^{i}$ of $d^{i}\in \mathbb{Z}$ and
$\chi _{\overrightarrow{A}}(\gamma _{\overrightarrow{\mu }})=\overset{L}{%
\underset{i=1}{\prod }}\chi _{A^{i}}(\gamma _{\mu ^{i}})$ for the character
$\chi _{A^{i}}$ of $Br_{d^{i}}$ labeled by $A^{i}$.
Expend the free energy
\begin{equation}
F_{CS}^{SO}(\mathcal{L};q,t;\overrightarrow{X})=\log Z_{CS}^{SO}(\mathcal{L}
;q,t;\overrightarrow{X})=\sum_{\overrightarrow{\mu }\neq \overrightarrow{0}
}F_{\overrightarrow{\mu }}^{SO}(\mathcal{L};q,t)p_{\overrightarrow{\mu }}(
\overrightarrow{X}),
\end{equation}
Then the reformulated invariants are defined by
\begin{equation}
g_{\overrightarrow{\mu }}(\mathcal{L};q,t)=\sum_{k|\overrightarrow{\mu }}%
\frac{\mu (k)}{k}F_{\overrightarrow{\mu }/k}(\mathcal{L};q^{k},t^{k}).
\end{equation}
The orthogonal LMOV conjecture assumes that \cite{C-C}
\begin{equation}
\frac{\mathfrak{z}_{\overrightarrow{\mu}}[1]^{2}}{2[\overrightarrow{\mu }]}
\left(g_{\overrightarrow{\mu }}
(\mathcal{L};q,t)-g_{\overrightarrow{\mu }}(\mathcal{L};q,-t)\right)
=\overset{\infty }{\sum_{g=0}}\sum_{\beta \in \mathbb{
Z}}N_{\overrightarrow{\mu },g,\beta }[1]_q^{g}\,t^{\beta },
\end{equation}
where $N_{\overrightarrow{\mu },g,\beta }$ are the integer coefficients and
vanish for sufficiently large $g$ and $|\beta |$.
This conjecture is a rigorous mathematical formulation of the LMOV type
conjecture about the colored Kauffman polynomial; while in \cite{BFM, M},
their conjecture emphasizes on the relationship between colored HOMFLY
and colored Kauffman. The integer coefficients $N_{\overrightarrow{\mu }%
,g,\beta }$ are closely related to the BPS numbers.
To derive an infinite-product formula, we will state the result
for a knot first, since the notations in the computation for a knot are
relatively simpler.

{\bf The case of a knot.}
By the orthogonal LMOV conjecture, $N_{B;g,\beta }$ vanish for sufficiently
large $g$ and $|\beta |$, thus $n_{B;\,g,\beta }$ vanish for sufficiently
large $g$ and $|\beta |$.
Finally the Chern-Simons partition function for orthogonal quantum group invariants
can be expressed as the following infinite-product formula
\begin{eqnarray}
\!\!\!\!\!\!\!\!\!\!\!\!\!\!\!\!\!\!
&&
\frac{\textsf{Z}_{CS}^{SO}(\mathcal{K};q,t;X)}{\textsf{Z}_{CS}^{SO}(\mathcal{K};q,-t;X)}
=
\underset{\mu \neq 0}{\prod }\overset{\infty }{\underset{g=0}{\prod }}
\underset{\beta \in \mathbb{Z}}{\prod }\underset{m=1}{\overset{\infty }{
\prod }}\underset{k=0}{\overset{g}{\prod }}\left( \frac{\big\langle%
1+q^{g-2k+2m}t^{\beta }X^{\mu }\big\rangle}{\big\langle1-q^{g-2k+2m}t^{\beta
}X^{\mu }\big\rangle}\right)^{\frac{mn_{\mu ,g,\beta }}{\mathfrak{z}_{\mu }}}
\nonumber \\
\!\!\!\!\!\!\!\!\!\!\!\!\!\!\!\!\!\!
= \!\!&&
\prod_{\mu \neq 0}\prod_{g=0}^\infty\prod_{\beta\in {\mathbb Z}}\prod_{k=0}^g\prod_{m=1}^\infty
\left(
\frac{{\mathcal R}(s= (2m+\Omega(q^{g-2k}t^\beta X^\mu;
\vartheta))(1-i\varrho(\vartheta))-1+i\sigma(\vartheta))}
{{\mathcal R}(s= (2m+\Omega(q^{g-2k}t^\beta X^\mu; \vartheta))(1-i\varrho(\vartheta))-1)}
\right)^{\frac{n_{\mu ,g,\beta }}{\mathfrak{z}_{\mu }}}\!\!\!\!\!\!\!.
\end{eqnarray}
\\

{\bf The case of a link.}
We generalize the symmetric product to the case of link as follows:
\begin{equation*}
\big\langle 1\pm \psi {\overrightarrow{X}}^\mu \big\rangle =
\prod_{i_{1,1},\ldots ,i_{1,\ell (\mu ^{1})},...,i_{L,1},\ldots
,i_{L,\ell (\mu ^{L})}}\Big(1\pm \psi \overset{L}{\prod_{\alpha =1}}\left(
(x_{i_{\alpha ,1}}^{\alpha })^{\mu _{1}^{\alpha }}\cdots (x_{i_{\alpha ,\ell
(\mu ^{\alpha })}}^{\alpha })^{\mu _{\ell (\mu ^{\alpha })}^{\alpha
}}\right) \Big)\,.
\end{equation*}
In a similar way the Chern-Simons partition function for orthogonal quantum group invariants
can be expressed as the following infinite-product formula
\begin{eqnarray}
\!\!\!\!\!\!\!\!
&&
\frac{\textsf{Z}_{CS}^{SO}(\mathcal{L};q,t;{\overrightarrow{X}})}
{\textsf{Z}_{CS}^{SO}(\mathcal{L
};q,-t;{\overrightarrow{X}})}=\underset{\overrightarrow{\mu }\neq
\overrightarrow{0}}{\prod }\overset{\infty }{\underset{g=0}{\prod }}\underset
{\beta \in \mathbb{Z}}{\prod }\underset{m=1}{\overset{\infty }{\prod }}
\underset{k=0}{\overset{g}{\prod }}\left( \frac{\big\langle%
1+q^{g-2k+2m}t^{\beta }(x^{1})^{\mu ^{1}}\cdot \cdot \cdot (x^{L})^{\mu ^{L}}
\big\rangle}{\big\langle1-q^{g-2k+2m}t^{\beta }(x^{1})^{\mu ^{1}}\cdot \cdot
\cdot (x^{L})^{\mu ^{L}}\big\rangle}\right) ^{\frac{mn_{\overrightarrow{\mu }
,g,\beta }}{\mathfrak{z}_{\overrightarrow{\mu }}}}
\nonumber \\
\!\!\!\!\!\!\!\! = \!\!\!\! &&
\underset{\overrightarrow{\mu }\neq
\overrightarrow{0}}{\prod }\overset{\infty }{\underset{g=0}{\prod }}\underset%
{\beta \in \mathbb{Z}}{\prod }\underset{m=1}{\overset{\infty }{\prod }}
\underset{k=0}{\overset{g}{\prod }}
\left(
\frac{{\mathcal R}(s= (2m+\Omega(q^{g-2k}t^\beta {\overrightarrow{X}}^\mu);
\vartheta))(1-i\varrho(\vartheta))-1+i\sigma(\vartheta))}
{{\mathcal R}(s= (2m+\Omega(q^{g-2k}t^\beta {\overrightarrow{X}}^\mu); \vartheta))
(1-i\varrho(\vartheta))-1)}\right)^{\frac{n_{\mu ,g,\beta }}{\mathfrak{z}_{\mu }}}\!\!\!\!\!\!\!.
\end{eqnarray}
\\

{\bf The case of the unknot.}
\begin{eqnarray}
\frac{\textsf{Z}_{CS}^{SO}(\bigcirc ;q,t;X)}{\textsf{Z}_{CS}^{SO}(\bigcirc ;q,-t;X)}
& = &
\underset{m=1}{\overset{\infty }{\prod }}\underset{i=-\infty }{\overset{\infty }{\prod
}}\left( \frac{(1+q^{2m}tx_{i})(1-q^{2m}t^{-1}x_{i})}{
(1-q^{2m}tz_{i})(1+q^{2m}t^{-1}x_{i})}\right) ^{m}
\nonumber \\
& = &
\prod_{i=-\infty}^{\infty}\prod_{m=1}^\infty \left(\frac{{\mathcal R}(s= (2m +
\Omega(tx_i; \vartheta))-1+i\sigma(\vartheta))}{{\mathcal R}(s= (2m +
\Omega(t^{-1}x_i;\vartheta))-1+i\sigma(\vartheta))}\right)
\nonumber \\
& \times &
\left(\frac{{\mathcal R}(s= (2m + \Omega(t^{-1}x_i; \vartheta))
-1)}{{\mathcal R}(s= (2m + \Omega(tx_i; \vartheta))-1)}\right)\,.
\end{eqnarray}

\section*{Acknowledgments}

We are indebted to Qingtao Chen for useful discussions at various stages of this work.
AAB would like to acknowledge the Conselho Nacional de Desenvolvimento Cient\'{i}fico e
Tecnol\'{o}gico (CNPq, Brazil) and Coordena\c{c}\~{a}o de Aperfei\c{c}amento de Pessoal
de N\'{i}vel Superior (CAPES, Brazil) for financial support.

\end{document}